\documentclass[twocolumn,aps,showpacs,prb,tightenlines,amsmath,amssymb]{revtex4}
\usepackage{graphicx}
\usepackage{amssymb}
\usepackage{dcolumn}
\usepackage{amsmath}
\usepackage{pifont}
\usepackage{bm}
\usepackage{colordvi}
\newcommand{\bgreek}[1]{\mbox{\boldmath$#1$\unboldmath}}

\begin{document}

\title{Electron spin relaxation in bilayer graphene}
\author{L. Wang}
\affiliation{Hefei National Laboratory for Physical Sciences at
Microscale and Department of Physics,
University of Science and Technology of China, Hefei,
Anhui, 230026, China}
\author{M. W. Wu}
\thanks{Author to whom correspondence should be addressed}
\email{mwwu@ustc.edu.cn.}
\affiliation{Hefei National Laboratory for Physical Sciences at
Microscale and Department of Physics, University of Science and
Technology of China, Hefei, Anhui, 230026, China}

\date{\today}

\begin{abstract}
Electron spin relaxation due to the D'yakonov-Perel' 
mechanism is investigated in bilayer graphene with only the lowest conduction 
band being relevant. The spin-orbit coupling is constructed 
from the symmetry group analysis with the parameters obtained by fitting to the 
numerical calculation according to the latest report by Konschuh {\it et al.} 
[Phys. Rev. B {\bf 85}, 115423 (2012)] 
from first principles. In contrast to single-layer graphene, the leading term of 
the out-of-plane component of the  spin-orbit coupling in 
bilayer graphene shows a Zeeman-like term with opposite effective 
magnetic fields in the two valleys. 
This Zeeman-like term opens a spin relaxation channel in the 
presence of intervalley scattering. It is shown that the intervalley 
electron-phonon scattering, which has not been reported in the previous literature, 
strongly suppresses the in-plane spin relaxation time at high temperature whereas 
the intervalley short-range scattering plays an important role in the in-plane 
spin relaxation especially at low temperature. A marked nonmonotonic dependence of the 
in-plane spin relaxation time on temperature with a minimum of several 
hundred picoseconds is predicted in the absence of the short-range scatterers. 
This minimum is comparable to the experimental data. 
Moreover, a peak in the electron density dependence of the
in-plane spin relaxation time at low temperature, which is very different from 
the one in semiconductors, is predicted. 
We also find a rapid decrease in the in-plane spin relaxation time 
with increasing initial spin polarization at low temperature, which 
is opposite to the situation in both semiconductors and single-layer 
graphene. A strong anisotropy between the out-of- and in-plane spin 
relaxations at high temperature is also revealed with the out-of-plane 
spin relaxation time being about two orders of magnitude larger than the in-plane one. 
Detailed comparisons of the temperature and electron 
density dependences of the spin relaxation with the existing experiments 
[Phys. Rev. Lett. {\bf 107}, 047206 (2011), Phys. Rev. Lett. {\bf 107}, 047207 (2011) 
and Nano Lett. {\bf 11}, 2363 (2011)] are reported.
 
\end{abstract}


\maketitle
\section{INTRODUCTION}
In recent years, single-layer graphene has received much attention due to its two dimensionality, Dirac-like energy spectrum and outstanding spin-coherence 
properties.\cite{novoselov,geim6,tombros448,fwang,beenakker,neto81,peres,abergel,
sarma,acik,geim83,goerbig,kotov,cooper} Specifically, the hyperfine interaction and
spin-orbit coupling (SOC) are considerably weak,\cite{kane,hernando74,min,yao,boettger,trauzettel,fischer,gmitra,
hernando103,abdelouahed,konschuh82} suggesting a long spin relaxation time (SRT) in single-layer graphene.
This long SRT makes single-layer graphene a promising candidate for spintronic application.
So far, a number of experiments on spin relaxation have been carried out, with the SRTs being
orders of magnitude shorter than theoretical prediction.\cite{tombros448,cho,jozsa100,tombros101,jozsa80,
popinciuc,han94,han102,shiraishi,pi,han105,yzhou,han107,avsar,han324,maassen,wojtaszek} This
pronounced discrepancy has provoked many investigations on the extrinsic 
effects of spin relaxation such as 
adatoms,\cite{abdelouahed,varykhalov,neto,ertler,dugaev,pzhang84,pzhang14} 
curvature,\cite{morozov,jeong,hernando74,pzhang112}
substrate effects\cite{ryu,dedkov,ertler} and 
contacts.\cite{popinciuc,han324,maassen,gallo}
Consisting of two layer graphene, bilayer graphene (BLG) 
also possesses good spin-coherence
properties.\cite{trauzettel,fischer,guinea1,konschuh85,mccann,korm}
Additionally, in contrast to single-layer graphene, BLG has a tunable gap caused by an out-of-plane electric
field,\cite{min2,castro,zhang,nilsson,konschuh85,mccann,mccann2} which can be
used to turn on or off the electronic transport. 
In addition, in the absence of space inversion symmetry, the intrinsic SOC in the pseudospin 
space of BLG provides an out-of-plane effective
magnetic field near the Dirac points whereas the intrinsic SOC of single-layer graphene only
induces a shift of the energy spectrum.\cite{ertler,yzhou,konschuh85} Moreover,
the electron-phonon coupling becomes more complex in BLG due to the
existence of more phonon modes.\cite{viljas,borysenko,cappelluti}
So far, some efforts have been made to investigate the
  intravalley electron-phonon coupling.\cite{viljas,borysenko,cappelluti}
Specifically, the intravalley electron-acoustic (AC) phonon coupling has been
  calculated by Viljas and Heikkil\"{a}\cite{viljas} with the continuum
  theory and also by Borysenko {\it et al.}\cite{borysenko} from first
  principles. To explore the intravalley electron-optical (OP) phonon coupling, the
  tight-binding model, density functional theory and first principles
  approach are employed by Viljas and Heikkil\"{a},\cite{viljas} Cappelluti {\it et
  al.}\cite{cappelluti} and Borysenko {\it et al.},\cite{borysenko}
respectively. However, the intervalley electron-phonon coupling, which
has been shown to play an important role in spin relaxation in rippled
single-layer graphene,\cite{pzhang112} has not been reported in the
previous literature. All these intriguing features 
make BLG an attractive material for investigation.

Very recently, some attention has been devoted to the spin relaxation 
in BLG.\cite{yang,han107,avsar,han324,diez,neumann}
Experimentally, the SRTs of BLG on SiO$_2$ substrate\cite{yang,han107,avsar} or in 
freely suspended BLG\cite{neumann} are reported to be of the order 
of $0.01$-$1\ $ns. Both Yang {\it et al.}\cite{yang} and Han and Kawakami\cite{han107}
measured the temperature and carrier dependences of the SRT. Yang {\it et al.}\cite{yang} 
showed that the SRT is weakly dependent
on temperature. However, Han and Kawakami\cite{han107} reported a marked
decrease of the SRT with the increase of the temperature. 
As for the carrier density dependence, the decrease of the SRT with
increasing carrier density at low temperature is reported in both experiments. 
However, at room temperature, Yang {\it et al.}\cite{yang} observed an 
increase of the SRT with increasing density whereas the SRT 
measured by Han and Kawakami\cite{han107} exhibits a marginal density dependence. 
In addition, Yang {\it et al.}\cite{yang} also showed that the SRT scales inversely
with the mobility at both room and low temperatures.
In contrast to the above two experiments based on the
micromechanically exfoliated BLG samples,
Avsar {\it et al.}\cite{avsar}
carried out the experiment in chemical vapor deposition grown BLG. The carrier density and
temperature dependences of the SRT measured are comparable to 
the ones by Yang {\it et al.}\cite{yang} both
quantitatively and qualitatively. Motivated by these experiments, Diez
and Burkard\cite{diez} calculated
the electron spin relaxation due to the D'yakonov-Perel' (DP) mechanism\cite{dp} with only the 
long-range electron-impurity
scattering included in BLG. They claimed to obtain a good agreement with the experimental 
data of Yang {\it et al.}.\cite{yang} It is noted that the effective 
SOC (referred to as the spin-orbit field in Ref.~\onlinecite{diez})
 used in their calculation 
is induced by an interlayer bias voltage together with the intrinsic SOC 
in the pseudospin space.\cite{guinea1} 
However, the intrinsic SOC in the pseudospin space 
becomes incomplete in the presence of the interlayer bias since the 
interlayer bias breaks the space inversion symmetry and
leads to additional extrinsic SOC 
terms in the pseudospin space.\cite{konschuh85} 
Also even with only the intrinsic SOC in the pseudospin space 
in their calculation, they overlooked 
the out-of-plane component of the effective SOC. 
As will be shown later, the leading term of this component 
leads to a Zeeman-like term in the 
two valleys and makes a dominant contribution 
to the in-plane spin relaxation. Moreover, they neglected the 
electron-phonon and electron-electron Coulomb scatterings. In fact, 
the electron-electron Coulomb scattering has been
demonstrated to be very important for the spin relaxation in semiconductor
systems\cite{wureview,wuning,weng1,glazov,jzhou,leyland,stich,lfhan43}
and also single-layer graphene at high temperature.\cite{yzhou} 
The electron-phonon scattering has also been shown to play an important role 
in spin relaxation in both semiconductors\cite{wureview} and single-layer 
graphene\cite{yzhou} at high temperature. Therefore, both have to be included 
in studying the spin relaxation.

In the present work, with the electron-electron Coulomb, 
(both the intra- and inter-valley) electron-phonon, possible 
short-range\cite{sarma} 
as well as long-range electron-impurity scatterings explicitly included, 
we investigate the electron spin relaxation 
due to the DP mechanism in BLG with only the lowest conduction 
band being relevant by the kinetic spin Bloch equation 
(KSBE) approach.\cite{wureview} 
The SOC of the lowest 
conduction band near the Dirac points ${\bf \Omega}^{\mu}({\bf k})$ is 
constructed from the symmetry group $C_3$ in the presence of an out-of-plane 
electric field.\cite{koster}
Its three components $\Omega_i^{\mu}({\bf k})$ $(i=x,y,z)$ can be 
written as (see Appendix~\ref{appA})
\begin{eqnarray}
 \Omega_x^{\mu}({\bf k})&=&\alpha_1(k)\sin\theta_{\bf k}+
\mu[\alpha_2(k)\sin 2\theta_{\bf k}\nonumber\\
&&\mbox{}+\alpha_3(k)\sin 4\theta_{\bf k}],\label{eq1}\\
\Omega_y^{\mu}({\bf k})&=&-\alpha_1(k)\cos\theta_{\bf k}+
\mu[\alpha_2(k)\cos 2\theta_{\bf k}\nonumber\\
&&\mbox{}-\alpha_3(k)\cos 4\theta_{\bf k}],\label{eq2}\\
\Omega_z^{\mu}({\bf k})&=&\mu\beta_1(k)+\beta_2(k)\cos 
3\theta_{\bf k},\label{eq3}
\end{eqnarray}
with $\mu=1(-1)$ standing for K(K$^{\prime}$) valleys and $k$ and $\theta_{\bf k}$ representing the magnitude and
the polar angle of the momentum ${\bf k}$ relative to K(K$^{\prime}$) points,
respectively. The coefficients $\alpha_i(k)$ ($i=1$-$3$) and
$\beta_{1,2}(k)$ are given in Appendix~\ref{appA} by fitting to
  the numerical results calculated according to the latest report 
by Konschuh {\it et al.}\cite{konschuh85} from first principles with both 
the intrinsic and extrinsic SOC terms in the pseudospin space included.
The out-of-plane component $\Omega_z^{\mu}({\bf k})$ [Eq.~(\ref{eq3})], 
which has not been reported in
the previous literature, is induced by the intrinsic SOC in the 
pseudospin space in the absence of space inversion symmetry,
whereas the in-plane component is contributed by both 
the intrinsic and extrinsic SOCs in the pseudospin space. 
This is very different from the case in single-layer 
graphene where the intrinsic SOC in the pseudospin space 
only induces a shift of the energy spectrum in the conduction 
band.\cite{ertler,yzhou,konschuh85}
We find that the magnitudes of both the out-of- and 
in-plane components decrease with the increase of $k$ at large momentum, indicating a
  suppression of the inhomogeneous broadening\cite{wuning} by increasing
momentum.\cite{konschuh85} This is very different 
from the case in both semiconductors
and single-layer graphene.\cite{wureview,yzhou}
Additionally, the leading term of
the out-of-plane component, i.e., $\mu\beta_1(k)$, serves as a Zeeman-like term
with opposite effective magnetic fields perpendicular to BLG plane in the 
two valleys, similar to the case in rippled
single-layer graphene.\cite{pzhang112} This Zeeman-like term, together with the
intervalley scattering, opens a spin relaxation channel. This
spin relaxation channel suppresses the in-plane SRT {\it significantly} 
at high temperature by the 
intervalley electron-phonon scattering and also 
at low temperature by the short-range scattering.\cite{sarma} In addition, we find that 
although such Zeeman-like terms also 
exist in the in-plane components of 
the SOC, their contributions to the
out-of-plane spin relaxation are marginal since these terms are of the high order 
of the momentum. This leads to a strong anisotropy between the 
in-plane and out-of-plane spin relaxations.
We also find that the in-plane SRT shows a marked 
nonmonotonic dependence on temperature with a minimum down to several hundred 
picoseconds in the absence of short-range scatterers. This minimum is comparable to 
the experimental data. The nonmonotonic behavior 
results from the crossover between the weak and strong intervalley electron-phonon 
scattering. Moreover, we predict a peak in the density 
dependence at low temperature, 
which is very different from the one in semiconductors.\cite{lfhan43,jiang} 
As for high temperature, the SRT shows a monotonic increase with increasing 
density. We also find that the in-plane SRT decreases 
rapidly with increasing initial 
spin polarization at low temperature, which is very different from the 
previous studies in both semiconductors\cite{stich,fzhang,korn,weng1} 
and single-layer graphene.\cite{yzhou} 
Finally, we report detailed comparisons of the temperature and electron 
density dependences with the existing experiments.\cite{yang,han107,avsar}

This paper is organized as follows. In Sec.~II, we present our model and construct the KSBEs.
The main results are given in Sec.~III. In Sec.~IIIA, 
we investigate the temperature, electron density and also initial spin polarization 
dependences of the electron spin relaxation in the 
absence of short-range scatterers. The anisotropy of the spin relaxation is 
also addressed in this part. In Sec.~IIIB, we show detailed comparisons of the 
temperature and electron density dependences of the spin relaxation with 
the existing experiments with the inclusion of short-range scatterers. 
We summarize in Sec.~IV.

\section{MODEL AND KSBEs}
We start our investigation from the AB-stacked BLG on SiO$_2$ substrate. The
spinless $\pi$-band structure can be described by the effective $4\times 4$ Hamiltonian:\cite{konschuh85}
\begin{eqnarray}
H_{\rm TB}(\tilde {\bf k})=\left(\begin{array}{cccc}
\Delta+V & \gamma_0 f(\tilde {\bf k}) & \gamma_4 f^*(\tilde {\bf k}) & \gamma_1 \\
\gamma_0 f^*(\tilde {\bf k}) & +V & \gamma_3 f(\tilde {\bf k}) & \gamma_4 f^*(\tilde {\bf k}) \\
\gamma_4 f(\tilde {\bf k}) & \gamma_3 f^*(\tilde {\bf k}) & -V & \gamma_0 f(\tilde {\bf k}) \\
\gamma_1 & \gamma_4 f(\tilde {\bf k}) & \gamma_0 f^*(\tilde {\bf k}) & \Delta-V
\end{array}\right)\label{eq4}
\end{eqnarray}
in the on-site orbital Bloch basis $\Psi_{\rm A_1}(\tilde {\bf k})$, $\Psi_{\rm B_1}(\tilde {\bf k})$,
$\Psi_{\rm A_2}(\tilde {\bf k})$ and
$\Psi_{\rm B_2}(\tilde {\bf k})$, where A$_1$ refers to the A sublattice in the lower layer, B$_2$
to the B sublattice in the upper layer, etc., and ${\tilde {\bf k}}$ represents the two-dimensional
wave vector counted from the $\Gamma$ point.
Here, $\gamma_0$ and $\gamma_1$ describe the nearest-neighbor intralayer and
interlayer hoppings whereas $\gamma_3$ and $\gamma_4$ stand for the indirect hoppings between two layers.
$\Delta$ represents the asymmetry in the energy shift of the on-site energy due to the interlayer
hopping. The potential $V=eE_zd_{\rm eff}/2$ with $e$, $E_z$ and $d_{\rm eff}$ representing the electron charge $(e>0)$,
the out-of-plane electric field and the effective electrostatic
bilayer distance, respectively. $f(\tilde {\bf k})=
e^{i\frac{a}{\sqrt{3}}{{\tilde k}_y}}[1+2e^{-i\frac{\sqrt{3}a}{2}{{\tilde k}_y}}
\cos(\frac{a}{2}{{\tilde k}_x})]$
is the nearest-neighbor structural function of the graphene hexagonal 
lattice with $a$ denoting the lattice constant. 
Near the Dirac points, to the first order of the momentum ${{\bf k}}$ 
originated from the K(K$^{\prime}$)
points, $f(\tilde {\bf k})\equiv p({\bf k})=-\sqrt{3}a(\mu {k}_x-i{k}_y)/2$.
Then the effective $4\times 4$ Hamiltonian near the Dirac points can be written as
\begin{eqnarray}
H^{\mu}({\bf k})=\left(\begin{array}{cccc}
\Delta+V & \gamma_0 p & \gamma_4 p^* & \gamma_1 \\
\gamma_0 p^* & +V & \gamma_3 p & \gamma_4 p^* \\
\gamma_4 p & \gamma_3 p^* & -V & \gamma_0 p \\
\gamma_1 & \gamma_4 p & \gamma_0 p^* & \Delta-V
\end{array}\right).\label{eq5}
\end{eqnarray}
By exactly diagonalizing this Hamiltonian, one obtains the eigenvalue $\epsilon_{\mu\nu {\bf k}}$ and the
eigenfunction $\psi^{\mu\nu}_{\bf k}$ with $\nu$ being the band index including two conduction bands and
two valence ones. In our calculation, we concentrate on the heavily $n$-doped case with only
the lowest conduction band being relevant whereas 
other bands are separated by large energy intervals. 
For convenience, the eigenvalue and the corresponding
eigenfunction of the lowest conduction band at ${\bf k}$ are denoted by $\epsilon_{\mu {\bf k}}$ and
$\psi^{\mu}_{\bf k}$, respectively.

By incorporating the spin degree of freedom, the effective Hamiltonian of the lowest
conduction band is given by
\begin{eqnarray}
H_{\rm eff}^{\mu}&=&\epsilon_{\mu{\bf k}}+{\bf \Omega}^{\mu}({\bf k})\cdot{\bgreek \sigma}/2.\label{eq6}
\end{eqnarray}
Here, ${\bgreek \sigma}$ are the Pauli matrices. ${\bf
  \Omega}^{\mu}({\bf k})$ stands for the effective magnetic field of the SOC, 
which is constructed from the symmetry analysis with 
the parameters obtained approximately 
by fitting to numerical results 
based on the report by Konschuh {\it et al.}.\cite{konschuh85} The analytical 
form of the SOC with three components $\Omega_i^{\mu}({\bf k})$ $(i=x,y,z)$ is 
shown in Eqs.~(\ref{eq1}-\ref{eq3}) and the explicit numerical 
calculation is given in Appendix~\ref{appA}. It is noted that at large momentum,
the analytical result agrees with the numerical one fairly well (see Appendix~\ref{appA}).

We then construct the microscopic KSBEs\cite{wureview} to study the electron spin relaxation in BLG. The
KSBEs read\cite{wureview}
\begin{eqnarray}
\partial_t\hat{\rho}_{\mu{\bf k}}=\partial_t\hat{\rho}_{\mu{\bf k}}|_{\rm coh}+\partial_t\hat{\rho}_{\mu{\bf k}}|_{\rm scat},
\label{eq7}
\end{eqnarray}
where $\hat{\rho}_{\mu{\bf k}}$ represent the density matrices of electrons
with the diagonal terms $\rho_{\mu{\bf k},\sigma\sigma}\equiv f_{\mu{\bf k}\sigma}\ (\sigma=\pm \frac{1}{2})$
describing the distribution functions and the off-diagonal ones $\rho_{\mu{\bf k},(1/2)(-1/2)}=\rho_{\mu{\bf k},(-1/2)(1/2)}^*$ denoting
the spin coherence. The coherent term is given by
\begin{eqnarray}
\partial_t\hat{\rho}_{\mu{\bf k}}|_{\rm coh}=-i[{\bf
  \Omega}^{\mu}({\bf k})\cdot{\bgreek \sigma}/2+\hat{\Sigma}^{\rm HF}_{\mu{\bf k}},
\hat{\rho}_{\mu{\bf k}}],
\label{eq8}
\end{eqnarray}
where $[A,B]\equiv AB-BA$ is the commutator; 
$\hat{\Sigma}^{\rm HF}_{\mu{\bf k}}=-\sum_{{\bf k}^{\prime}}V_{{\bf k},{\bf k}^{\prime}}^{\mu}
I^{\mu}_{{\bf k}{\bf k}^{\prime}}\hat{\rho}_{\mu{\bf k}^{\prime}}$ represents the 
Coulomb Hartree-Fock (HF) term.\cite{weng1}
$V_{{\bf k},{\bf k}^{\prime}}^{\mu}$ denotes the screened Coulomb potential with its form
given in Appendix~\ref{appB}. The form factor
$I^{\mu}_{{\bf k}{\bf k}^{\prime}}=|{\psi_{\bf k}^{\mu}}^{\dagger}\psi_{{\bf k}^{\prime}}^{\mu}|^2$.
$\partial_t\hat{\rho}_{\mu{\bf k}}|_{\rm scat}$ are the scattering terms including the electron-electron Coulomb
($|V_{{\bf k},{\bf k}^{\prime}}^{\mu}|^2$), long-range 
electron-impurity ($|U_{{\bf k},{\bf k}^{\prime}}^{\mu}|^2$), intravalley
electron-AC phonon ($|M^{\rm AC}_{{\mu{\bf k}},{\mu^{\prime}{\bf
      k}^{\prime}}}|^2$), electron-OP phonon
($|M^{\rm OP}_{{\mu{\bf k}},{\mu^{\prime}{\bf k}^{\prime}}}|^2$), electron-remote-interfacial (RI) phonon
($|M^{\rm RI}_{{\mu{\bf k}},{\mu^{\prime}{\bf k}^{\prime}}}|^2$) and especially the intervalley electron-phonon scattering
including the electron-${\rm KA_1^{\prime}}$ phonon ($|M^{\rm KA_1^{\prime}}_{{\mu{\bf k}},{\mu^{\prime}{\bf k}^{\prime}}}|^2$),
electron-${\rm KA_2^{\prime}}$ phonon ($|M^{\rm KA_2^{\prime}}_{{\mu{\bf k}},{\mu^{\prime}{\bf k}^{\prime}}}|^2$),
electron-${\rm KE^{\prime}}$ phonon ($|M^{\rm KE^{\prime}}_{{\mu{\bf k}},{\mu^{\prime}{\bf k}^{\prime}}}|^2$) and
electron-${\rm KZ}$ phonon ($|M^{\rm KZ}_{{\mu{\bf k}},{\mu^{\prime}{\bf k}^{\prime}}}|^2$) scatterings.
Their detailed expressions can be found in Ref.~\onlinecite{yzhou}.\cite{delta} 
Here, ${\rm KA_1^{\prime}}$, ${\rm KA_2^{\prime}}$ and ${\rm KE^{\prime}}$
are the in-plane phonon modes corresponding to the representations $A_1^{\prime}$, $A_2^{\prime}$ and $E^{\prime}$ of group $D_{3h}$,
respectively whereas ${\rm KZ}$ stands for the out-of-plane 
phonon mode.\cite{piscanec,lazzeri,borysenko,rana} It is
noted that the intravalley electron-phonon scattering has been reported in the 
previous literature as mentioned in the introduction.\cite{viljas,borysenko}
However, to the best of our knowledge, there is no specific report 
to the intervalley electron-phonon scattering.
In this work, we derive the intervalley electron-phonon scattering matrix elements
using the tight-binding model according to the arXiv version of Ref.~\onlinecite{viljas}. 
The scattering matrix elements are given by
\begin{eqnarray}
&&\hspace{-0.4cm}|M^{\rm KA_1^{\prime}}_{{\mu{\bf k}},{\mu^{\prime}{\bf k}^{\prime}}}|^2=\frac{3\hbar^2}{4\rho\Omega_{\rm KA_1^{\prime}}}\delta_{\mu^{\prime},-\mu}
\Big[|{\psi^{\mu}_{{\bf k}}}^{\dagger}(2\sqrt{3}\gamma_0^{\prime}\sigma^{23}_{\rm D}+a\gamma_4^{\prime}\gamma^0_{\rm D}/l_4)\nonumber\\
&&\mbox{}\hspace{0.1cm}\times\psi^{\mu^{\prime}}_{{\bf k}^{\prime}}|^2+|{\psi^{\mu}_{{\bf k}}}^{\dagger}
(2\sqrt{3}i\gamma_0^{\prime}\sigma^{01}_{\rm D}
-a\gamma_4^{\prime}\gamma^3_{\rm D}\gamma^5_{\rm D}/l_4)\psi^{\mu^{\prime}}_{{\bf k}^{\prime}}|^2\Big],\label{eq9}\\
&&\hspace{-0.4cm}|M^{\rm KA_2^{\prime}}_{{\mu{\bf k}},{\mu^{\prime}{\bf k}^{\prime}}}|^2=\frac{3a^2{\gamma_4^{\prime}}^2}{2\rho l_4^2}
\frac{\hbar^2}{2\Omega_{\rm KA_2^{\prime}}}\delta_{\mu^{\prime},-\mu}
\Big[|{\psi^{\mu}_{{\bf k}}}^{\dagger}\gamma^3_{\rm D}\gamma^5_{\rm D}\psi^{\mu^{\prime}}_{{\bf k}^{\prime}}|^2\nonumber\\
&&\mbox{}\hspace{0.1cm}+|{\psi^{\mu}_{{\bf k}}}^{\dagger}\gamma^0_{\rm D}\psi^{\mu^{\prime}}_{{\bf k}^{\prime}}|^2\Big],\label{eq10}\\
&&\hspace{-0.4cm}|M^{\rm KE^{\prime}}_{{\mu{\bf k}},{\mu^{\prime}{\bf k}^{\prime}}}|^2=\frac{3a^2}{\rho l_4^2}
\frac{\hbar^2}{4\Omega_{\rm KE^{\prime}}}\delta_{\mu^{\prime},-\mu}
\Big[|{\psi^{\mu}_{{\bf k}}}^{\dagger}(\gamma_3^{\prime}\gamma^1_{\rm D}\gamma^5_{\rm D}-i\gamma_3^{\prime}\gamma^2_{\rm D}\nonumber\\
&&\mbox{}\hspace{0.1cm}-\gamma_4^{\prime}\gamma^0_{\rm D})\psi^{\mu^{\prime}}_{{\bf k}^{\prime}}|^2+|{\psi^{\mu}_{{\bf k}}}^{\dagger}
(i\gamma_3^{\prime}\gamma^2_{\rm D}-\gamma_3^{\prime}\gamma^1_{\rm D}\gamma^5_{\rm D}-\gamma_4^{\prime}\gamma^0_{\rm D})
\psi^{\mu^{\prime}}_{{\bf k}^{\prime}}|^2\nonumber\\
&&\mbox{}\hspace{0.1cm}+4|{\psi^{\mu}_{{\bf k}}}^{\dagger}\gamma_4^{\prime}\gamma^3_{\rm D}\gamma^5_{\rm D}
\psi^{\mu^{\prime}}_{{\bf k}^{\prime}}|^2\Big],\label{eq11}\\
&&\hspace{-0.4cm}|M^{\rm KZ}_{{\mu{\bf k}},{\mu^{\prime}{\bf k}^{\prime}}}|^2=\frac{{\gamma_1^{\prime}}^2}{\rho}
\frac{\hbar^2}{2\Omega_{\rm KZ}}\delta_{\mu^{\prime},-\mu}
|{\psi^{\mu}_{{\bf k}}}^{\dagger}(\gamma^1_{\rm D}\gamma^5_{\rm D}+i\gamma^2_{\rm D})\psi^{\mu^{\prime}}_{{\bf k}^{\prime}}|^2,\nonumber\\
\label{eq12}
\end{eqnarray}
where $\rho$ is the mass density of the BLG;\cite{viljas} $\Omega_{\rm KA_1^{\prime}}$, $\Omega_{\rm KA_2^{\prime}}$,
$\Omega_{\rm KE^{\prime}}$ and $\Omega_{\rm KZ}$ are the energy spectra of ${\rm KA_1^{\prime}}$, ${\rm KA_2^{\prime}}$,
${\rm KE^{\prime}}$, and ${\rm KZ}$ phonon modes, respectively;\cite{piscanec,lazzeri,borysenko}
$\gamma_i^{\prime}$ ($i=0,1,3,4$) is the derivative of $\gamma_i$
with respect to the corresponding hopping bond length;\cite{cappelluti} $l_4$ is the bond length of the interlayer hopping
$\gamma_4$; $\gamma^i_{\rm D}$ ($i=0$-$3$, $5$), $\sigma^{01}_{\rm D}$ and $\sigma^{23}_{\rm D}$
are $4\times 4$ Dirac matrices given in Appendix~\ref{appC}.\cite{peskin} We will show that the above intervalley
electron-phonon scattering plays a significant
role in the in-plane spin relaxation in our investigation. The
remaining scattering matrix elements such as $|V_{{\bf k},{\bf k}^{\prime}}^{\mu}|^2$, $|U_{{\bf k},{\bf k}^{\prime}}^{\mu}|^2$,
and $|M^{\lambda}_{{\mu{\bf k}},{\mu^{\prime}{\bf k}^{\prime}}}|^2$
($\lambda={\rm AC,\ OP\ and\ RI}$) are given in Appendix~\ref{appB}. 
It is noted that we also include the short-range scattering 
in our calculation with the scattering term laid out later in this paper.

\begin{table}[htb]
\caption{Parameters used in the calculation (from Ref.~\onlinecite{konschuh85} unless otherwise specified).}
\begin{tabular}{llll}
  \hline
  \hline
  $\Delta$ &$9.7$ meV  &$\gamma_0$   &$2.6$ eV\\
  $\gamma_1$   &$0.339$ eV &$\gamma_3$ &$0.28$ eV\\
  $\gamma_4$   &$-0.14$ eV  &$\lambda_{{\rm I}1}$   &$12$ $\mu$eV\\
  $\lambda_{{\rm I}2}$ &$10$ $\mu$eV &$\lambda_0$ &$(5+10 E^*_z)$ $\mu$eV\\
  $\lambda^{\prime}_0$ &$(5-10 E^*_z)$ $\mu$eV &$\lambda_3$ &$1.5 E^*_z$ $\mu$eV\\
  $\lambda_4$ &$(-12-3 E^*_z)$ $\mu$eV &$\lambda_4^{\prime}$ &$(-12+3 E^*_z)$ $\mu$eV\\
  $a$ &$2.46$ \r{A} &$d_{\rm eff}$ &$0.1$ nm\\
  $d$ &$0.4$ nm $^a$ &$r_s$ &$0.8$ $^a$\\
  $v_{\rm F}$   &$8\times 10^5$ m/s $^b$ &$v_{\rm ph}$ &$2\times 10^4$
  m/s $^a$\\
  $D_{\rm AC}$  &$15$ eV $^c$ &$\rho$ &$1.52\times 10^{-7}$
  g/cm$^{2}$ $^d$\\
  $\omega_1^{\rm RI}$  &$59$ meV $^a$ &$\omega_2^{\rm RI}$ &$155$ meV $^a$\\
  $g_1$  &$5.4\times 10^{-3}$ $^a$ &$g_2$ &$3.5\times 10^{-2}$ $^a$\\
  $\Omega_{\rm LT}$ &$196$ meV $^a$ &$\Omega_{\rm ZO}$  &$104$ meV $^c$\\
  $\gamma_0^{\prime}$ &$4.4$ eV/\r{A} $^e$ &$\gamma_1^{\prime}$
  &$0.61$ eV/\r{A} $^e$\\
  $\gamma_3^{\prime}$ &$0.54$ eV/\r{A} $^e$ &$\gamma_4^{\prime}$
  &$0.3$ eV/\r{A} $^e$\\
  $\Omega_{\rm KA_1^{\prime}}$  &$161.2$ meV &$\Omega_{\rm
    KA_2^{\prime}}$ &$125$ meV $^c$\\
  $\Omega_{\rm KE^{\prime}}$   &$152$ meV $^c$ &$\Omega_{\rm KZ}$
  &$65$ meV $^c$\\
  $l_3$ &$3.64$ \r{A} &$l_4$ &$3.64$ \r{A}\\
  $Z_i$ &$1$\\
\hline
\hline
\end{tabular}
\begin{tabular}{llll}
\hspace{-1cm}$^{\rm a}$Reference \onlinecite{yzhou}. & & &\\
\hspace{-1cm}$^{\rm b}$Reference \onlinecite{divincenzo}. & & &\\
\hspace{-1cm}$^{\rm c}$Reference \onlinecite{borysenko}. & & &\\
\hspace{-1cm}$^{\rm d}$Reference \onlinecite{viljas}. & & &\\
\hspace{-1cm}$^{\rm e}$Reference \onlinecite{cappelluti}. & & &\\
\hspace{-1.cm} It is noted that $E^*_z$ is the magnitude of $E_z$ in the \\
\hspace{-1.cm} unit of V/nm.
\end{tabular}
\end{table}

\section{NUMERICAL RESULTS}
By numerically solving the KSBEs, one obtains the time evolution of
spin polarization along direction ${\bf n}$
$P(t)=\sum_{\mu{\bf k}}{\rm Tr}[\hat{\rho}_{\mu{\bf k}}(t){\bgreek \sigma}\cdot{\bf n}]/N_e$
with $N_e$ being the electron density. Then the SRT can be determined
from the slope of the envelope of the spin polarization $P(t)$.\cite{wureview}
The initial spin polarization $P(0)$ is set to be $2.5$~\% and the spin-polarization
direction ${\bf n}$ is along the $x$-axis unless otherwise specified. 
It is noted that all the parameters used in our calculation are 
listed in Table~I. 
It is also noted that the SOC is calculated both fully numerically following
the approach by Konschuh {\em et al.}\cite{konschuh85} and 
analytically by using Eqs.~(\ref{eq1}-\ref{eq3}). 
The SRT calculated 
with the analytical form of the SOC and with the explicit 
numerical one agrees fairly well with each other.

\subsection{Spin relaxation in BLG with high mobilities}
In our calculation, the out-of-plane electric field is chosen to be a typical 
value $E_z=0.14\ $V/nm\cite{han107} and
the long-range impurity density is taken to be $N_i=1.5\times 10^{11}\
$cm$^{-2}$. The corresponding mobility in our investigation
is of the order of $10^4\ $cm$^2$/(V\,s), which
is about $1$-$2$ orders of magnitude larger than those reported in the 
existing experiments in BLG on 
SiO$_2$ substrate\cite{yang,han107,avsar} and comparable to the ones in the experiments in 
freely suspended BLG.\cite{neumann} 
It is noted that the short-range scattering is not included in 
this subsection.

\begin{figure}[bth]
\centering
\includegraphics[width=8.5cm]{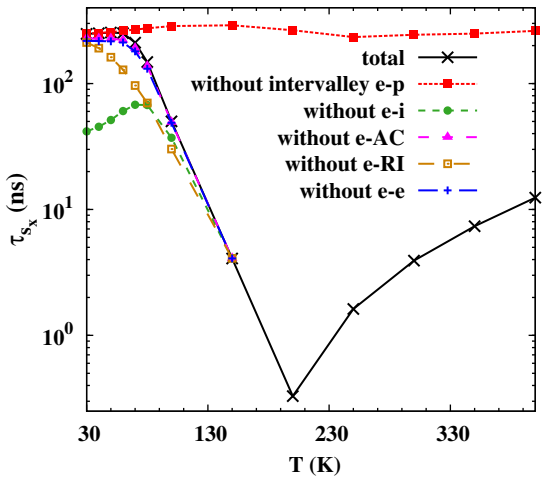}
\caption{(Color online) Total SRT along the $x$-axis $\tau_{s_x}$ ($\times$)
and the calculation without the long-range electron-impurity ($\bullet$),
electron-AC-phonon ($\blacktriangle$), electron-RI-phonon ($\square$), 
electron-electron Coulomb ($+$)
or intervalley electron-phonon ($\blacksquare$) scattering as function of temperature $T$.
It is noted that when $T> 150$~K, only the total SRT 
and the calculation without the intervalley 
electron-phonon scattering are shown since the remaining ones almost coincide with the total SRT. In the calculation, the electron 
density $N_e=3\times 10^{12}\ $cm$^{-2}$.}
\label{fig1}
\end{figure}

\subsubsection{Temperature dependence of spin relaxation}
We first investigate the temperature dependence of the spin relaxation.
In Fig.~\ref{fig1}, the SRT along the $x$-axis $\tau_{s_x}$ is plotted as function of temperature $T$.
It is seen that $\tau_{s_x}$ changes little with the temperature
at low temperature (i.e., $T\le 60\ $K). However, as $T$ becomes larger,
the SRT decreases dramatically with increasing temperature.
In particular, the SRT reaches down to several hundred picoseconds at $T=200\ $K, which is comparable to the
experimental values.\cite{yang,han107,avsar,neumann}
A detailed comparison with the existing experiments will be discussed 
in Sec.~IIIB. When $T$ further increases, the SRT begins to increase with $T$.

To understand the behavior of the spin relaxation, we show in Fig.~\ref{fig1}
the temperature dependence of the SRT calculated
with the electron-electron Coulomb, electron-RI-phonon, electron-AC-phonon,
long-range electron-impurity, or intervalley electron-phonon scattering
removed, separately. We find that at low temperature, the
intervalley electron-phonon scattering is negligible. Therefore, spins
relax independently in the two valleys and the SRT is solely determined by the
intravalley spin relaxation channel.
Specifically, the long-range electron-impurity scattering is dominant,
which leads to a weak temperature dependence of the SRT.\cite{yzhou,pzhang112}
With the increase of the temperature, 
the contribution of the long-range electron-impurity scattering becomes
marginal whereas the intervalley electron-phonon scattering
becomes important. This intervalley scattering opens another spin relaxation channel
together with the Zeeman-like term with opposite effective magnetic
fields in the two valleys, which is similar to the
case in rippled single-layer graphene.\cite{pzhang112} This
Zeeman-like term is given by $\mu\beta_1(k)$ [Eq.~(\ref{eq3})]
approximately. Then the spin relaxation due
to the above intervalley spin relaxation channel 
can be approximated by the rate equations\cite{pzhang112}
\begin{eqnarray}
\dot{{\bf S}}_{\mu k}(t)+{\bf S}_{\mu k}(t)\times {\bgreek \omega}_{\mu k}+
\frac{{\bf S}_{\mu k}(t)-{\bf S}_{-\mu k}(t)}{\tau_v(k)}=0,\label{eq13}
\end{eqnarray}
where ${\bf S}_{\mu k}$, ${\bgreek \omega}_{\mu k}=[\mu\beta_1(k)]\hat{{\bf z}}$
and $\tau_v(k)$ represent the in-plane spin vectors in each valley, spin precession vector along the $z$-axis and
the intervally electron-phonon scattering time, respectively.
According to the report by Zhang {\it et al.},\cite{pzhang112}
the in-plane SRT is given by
\begin{eqnarray}
\tau_{s_{x,y}}(k)=\left\{
\begin{array}{ll}
\tau_v(k) \mbox{{\rm \ \ weak\ scattering}}\\
{\rm \hspace{0.9cm}limit}\ (|\beta_1(k)|\tau_v(k)\ge 1)\\
\frac{2}{|\beta_1(k)|^2\tau_v(k)} \mbox{{\rm \ \ \ strong scattering}}\\
{\rm \hspace{0.9cm}limit}\ (|\beta_1(k)|\tau_v(k)\ll 1)
\end{array}\right.\label{eq14}
\end{eqnarray}
Here, $\tau_{s_y}$ stands for the SRT along the $y$-axis.
Since the system we investigate is always in the 
degenerate regime, we take $\tau_{s_{x,y}}(k)\approx \tau_{s_{x,y}}(k_F)$ 
approximately  with $k_F$ representing the Fermi wave vector.
In our calculation, the SRT along the $x$-axis is solely
 determined by the intervalley electron-phonon
scattering at $T\sim 100$-$400\ $K. At $T=100\ $K, $|\beta_1(k_F)|\tau_v(k_F)\approx 160$, i.e.,
the intervalley electron-phonon scattering
is in the weak scattering limit. As a result, the SRT $\tau_{s_x}=\tau_v(k_F)$ according to Eq.~(\ref{eq14}),
which decreases with the enhancement of the intervalley scattering as the temperature
increases. Nevertheless, at $T=400\ $K, $|\beta_1(k_F)|\tau_v(k_F)\approx 0.04$, i.e.,
the intervalley scattering is in the strong scattering limit. Then the SRT
$\tau_{s_x}=2/[|\beta_1(k_F)|^2\tau_v(k_F)]$ from Eq.~(\ref{eq14}), which increases with increasing of $T$.
Therefore, the SRT first decreases then increases with increasing temperature when
the intervalley scattering changes from the weak to strong scattering limit. The crossover
between the weak and strong intervalley scattering limit is determined
by $\tau_v(k_F)^{-1}\approx |\beta_1(k_F)|$.\cite{pzhang112} At this crossover point,
\begin{eqnarray}
\tau_{s_x}\approx |\beta_1(k_F)|^{-1}.\label{eq15}
\end{eqnarray}
Here $\tau_{s_x}\approx 315\ $ps,
very close to the value $330\ $ps shown in the figure. It is noted that the electron-electron Coulomb scattering
is unimportant in the whole temperature regime.

\subsubsection{Electron-density dependence of spin relaxation}
\begin{figure}[bth]
\centering
\includegraphics[width=8.5cm]{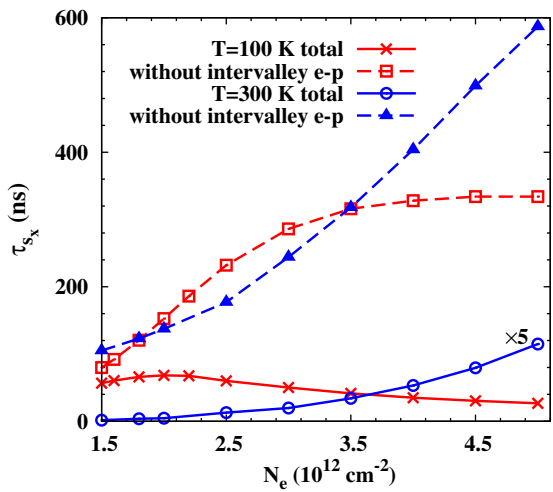}
\caption{(Color online) SRT along the $x$-axis $\tau_{s_x}$ as function of the electron density $N_e$
with $\times$ ($\circ$) and without $\square$ ($\blacktriangle$)
the intervalley electron-phonon scattering at $T=100\ (300)\ $K.}
\label{fig2}
\end{figure}

Then we turn to study the electron-density dependence. In Fig.~\ref{fig2}, the SRT along the $x$-axis
is plotted against the electron density at $T=100$ $(300)\ $K. One observes a peak
in the electron-density dependence of the SRT at $T=100\ $K whereas the SRT increases with the increase of
the electron density when $T=300\ $K. We first focus on the case at $T=100\ $K and
show the SRT with the intervalley electron-phonon scattering excluded in Fig.~\ref{fig2}.
It is seen that when the electron density is low, the intervalley electron-phonon scattering is marginal.
The SRT is then determined by the intravalley spin relaxation channel and hence 
increases with the suppression of the inhomogeneous broadening\cite{wuning} 
as the electron density increases.\cite{konschuh85}
The suppression of the inhomogeneous broadening can be understood from
the decrease of the spin splitting with increasing momentum at high
momentum [see Eqs.~(\ref{eq1}-\ref{eq3})], as shown in Fig.~\ref{fig9} in Appendix~\ref{appA}. This
behavior is very different from the case in both semiconductors and
single-layer graphene.\cite{wureview,yzhou}
When the electron density further increases, the intervalley electron-phonon scattering becomes important.
This intervalley scattering is always in the weak scattering limit with
$|\beta_1(k_F)|\tau_v(k_F)\ge 40$ and therefore $\tau_{s_x}=\tau_v(k_F)$ according to Eq.~(\ref{eq14}).
As a result, the SRT decreases with the increase of the intervalley scattering
and hence the increase of the electron density.\cite{pzhang112} Therefore, in
the whole electron-density regime, a peak is observed. It is noted that 
this peak is very different from the one predicted by Jiang and Wu\cite{jiang} 
in semiconductors where the peak is attributed to the crossover from 
the nondegenerate-to-degenerate limit. As for the case at $T=300\ $K,
the SRT is always dominated by the intervalley electron-phonon scattering 
as shown in Fig.~\ref{fig2} by comparing the SRT calculated with and without 
the intervalley electron-phonon scattering.
Since this intervalley scattering is in
the strong scattering limit with $|\beta_1(k_F)|\tau_v(k_F)\le 0.75$, the SRT
$\tau_{s_x}=2/[|\beta_1(k_F)|^2\tau_v(k_F)]$ from Eq.~(\ref{eq14}). This SRT
increases with increasing electron density since both
$|\beta_1(k_F)|$ and the intervalley scattering time $\tau_v(k_F)$ decrease with the
increase of the electron density.\cite{pzhang112,konschuh85}

\begin{figure}[bth]
\centering
\includegraphics[width=8.5cm]{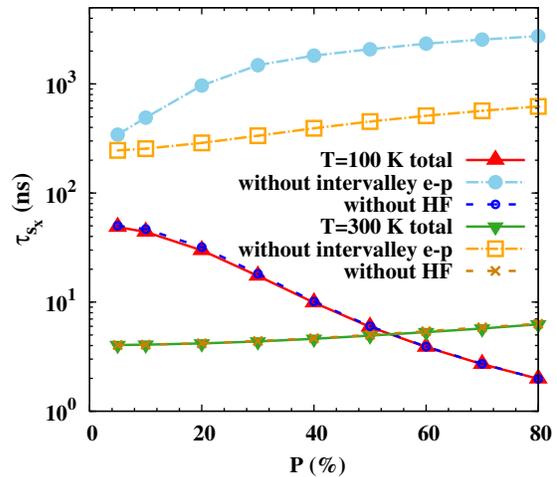}
\caption{(Color online) Total SRT along the $x$-axis $\tau_{s_x}$ $\blacktriangle$ ($\blacktriangledown$)
and the calculation without the Coulomb HF term $\circ$ ($\times$) or
the intervalley electron-phonon scattering $\bullet$
($\square$) as function of the initial spin polarization $P$ at $T=100\ (300)\ $K. 
In the calculation, the electron density $N_e=3\times 10^{12}$cm$^{-2}$.}
\label{fig3}
\end{figure}

\subsubsection{Influence of initial spin polarization}
The initial spin polarization dependences of the SRT along the $x$-axis at $T=100$ and $300\ $K are
shown in Fig.~\ref{fig3}. It is seen that the SRT decreases rapidly with the
increase of initial spin polarization $P$ at $T=100\ $K whereas the SRT shows a slight increase
with increasing $P$ at $T=300\ $K. The rapid decrease of the SRT at $T=100\ $K is
very different from the previous studies in both 
semiconductors\cite{stich,fzhang,korn,weng1} and
single-layer graphene\cite{yzhou} where the SRT increases significantly with
increasing initial spin polarization. In
the previous studies, the increase of the SRT originates from the
contribution of the Coulomb HF term.\cite{weng1} This term serves as an
effective magnetic field along the direction of the spin
polarization, which blocks the spin precession induced by the Dresselhaus\cite{dresselhaus} 
and/or Rashba\cite{rashba} SOCs and slows down the spin relaxation. 
However, the contribution of the Coulomb HF
term is marginal to the spin relaxation in our investigation,
which is shown in Fig.~\ref{fig3} by comparing the the SRT calculated with and
without the Coulomb HF term. Here, the SRT is determined by the intervalley electron-phonon
scattering, as shown in Fig.~\ref{fig3} by comparing the calculation
with and without the intervalley electron-phonon scattering. The intervalley 
electron-phonon scattering transfers electrons between the two valleys. The transferred 
electrons experience opposite effective magnetic fields induced by the Zeeman-like 
term in the two valleys, which are not affected by the effective magnetic field 
induced by the Coulomb HF term. Moreover, the 
intervalley electron-phonon scattering is in the weak scattering limit at 
low temperature and hence the SRT $\tau_{s_x}=\tau_v(k_F)$ [see Eq.~(\ref{eq14})].
The intervalley scattering time $\tau_v(k_F)$ decreases
significantly with the increase of the initial spin polarization as the density of states increases with the
polarization. Therefore, the SRT shows a rapid decrease with the initial spin polarization at $T=100\ $K.
As for the case of $T=300\ $K, the SRT is also determined by the intervalley electron-phonon scattering
in comparison with the case of $T=100\ $K. However, in contrast to the case of $T=100\ $K,
the intervalley scattering is in the strong scattering limit. 
This makes the SRT present an opposite
trend to the case of $T=100\ $K. A mild increase of the SRT results from
the insensitivity of $\tau_v(k_F)$ to the initial spin polarization at $T=300\ $K.

\begin{figure}[bth]
\centering
\includegraphics[width=8.5cm]{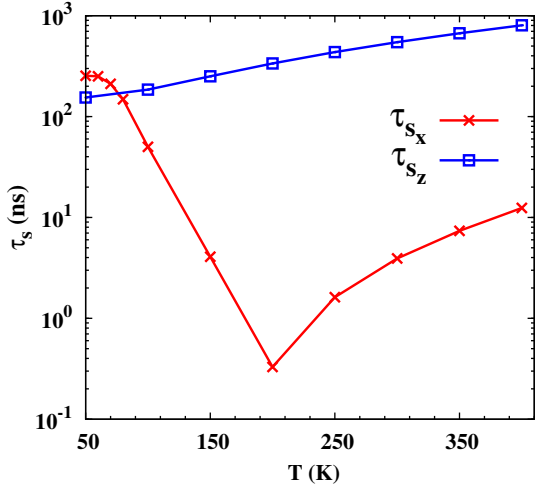}
\caption{(Color online) SRT along the $x$ $(z)$-axis $\tau_{s_x}$ $(\tau_{s_z})$ as function of temperature $T$. In the calculation, the electron density $N_e=3\times 10^{12}\ $cm$^{-2}$.}
\label{fig4}
\end{figure}

\subsubsection{Anisotropy of spin relaxation}
We also address the anisotropy of the spin relaxation 
with respect to the spin polarization direction.
Due to the existence of the out-of-plane effective magnetic field [see Eq.~(\ref{eq3})],
the in-plane SRTs are identical.\cite{pzhang112} For comparison,
we plot the temperature dependence of the SRT along the $x$ ($z$)-axis
as the red curve with crosses (blue curve with open squares) in Fig.~\ref{fig4}.
It is seen that in comparison with the SRT along the $z$-axis, 
the one along the $x$-axis is comparable at low temperature and 
orders of magnitude smaller at high temperature. 
As mentioned previously, the SRT along the
$x$-axis is strongly suppressed by the intervalley spin 
relaxation channel induced by the intervalley electron-phonon
scattering and the Zeeman-like term in the two valleys at high temperature.
However, the contribution of this spin relaxation channel to the SRT
along the $z$-axis is marginal [see Eqs.~(\ref{eq1}-\ref{eq2})].
Therefore, a strong anisotropy between the out-of- and in-plane spin relaxations is observed at high temperature.

\begin{figure}[tpb]
  \includegraphics[width=8.8cm]{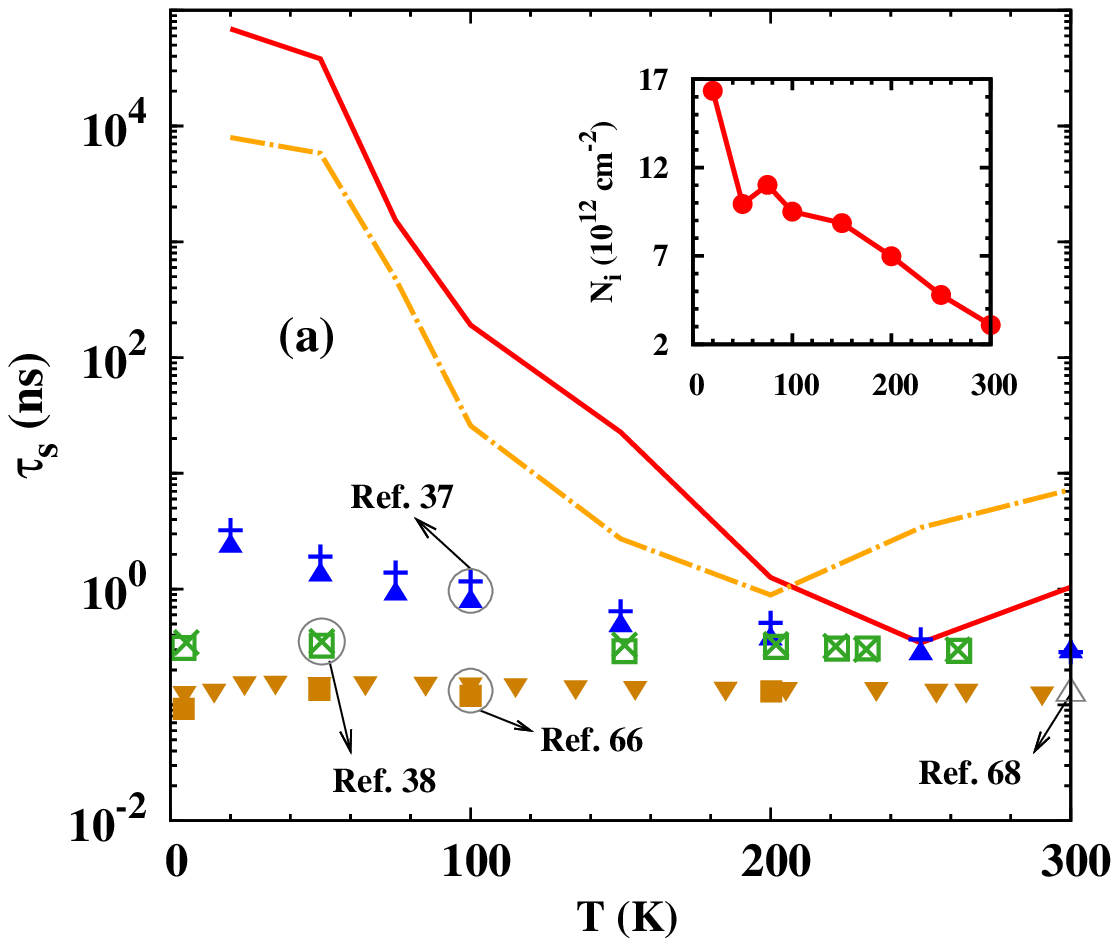}
  \includegraphics[width=9.5cm]{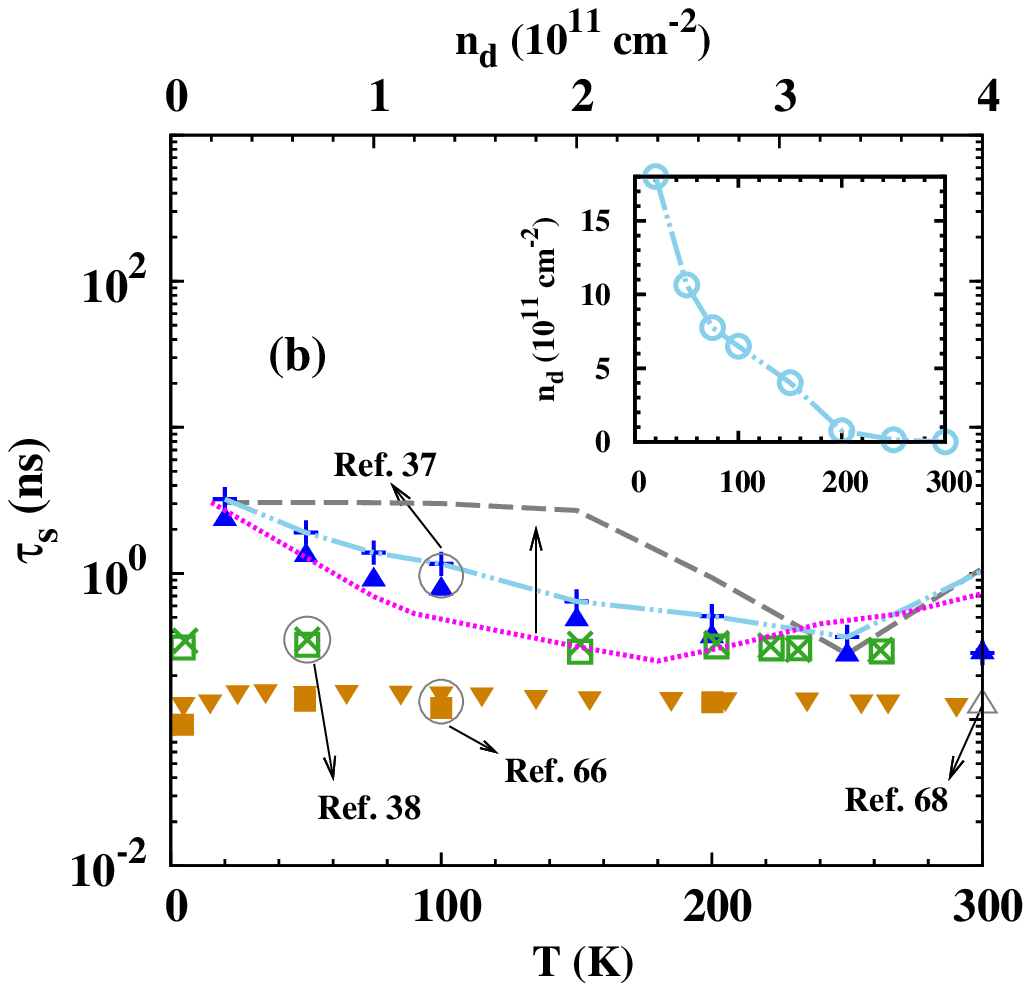}
\caption{(Color online) In-plane SRT $\tau_s$ as function of temperature $T$. 
$+$ ($\blacktriangle$) corresponds to
the case of the gate voltage V$_{\rm CNP}+20\ (60)\ $V in the experiment 
of Han and Kawakami;\cite{han107} $\times$ ($\square$) stands for the 
case of electron density $N_e=0.75\ (1.5)\times 10^{12}\ $cm$^{-2}$ in the experiment
of Avsar {\it et al.};\cite{avsar} $\blacktriangledown$ ($\blacksquare$) represents the case of
$N_e=0.7\ (2.2)\times 10^{12}\ $cm$^{-2}$ in the experiment of
 Yang {\it et al.};\cite{yang} $\triangle$ denotes 
the case of $N_e=10^{11}\ $cm$^{-2}$ 
in the experiment of Neumann {\it et al.}.\cite{neumann}
(a) Red solid (orange chain) curve
stands for our calculation corresponding to the case of
 V$_{\rm CNP}+20\ (60)\ $V 
in the absence of the short-range scatterers. In the inset, we show the 
temperature dependence of the 
long-range impurity density $N_i$ 
 corresponding to the case of V$_{\rm CNP}+20\ $V. 
(b) Grey dashed and skyblue double-dotted chain 
curves are obtained from
 our calculations corresponding to the case of V$_{\rm CNP}+20\ $V 
by including the short-range scatterers,
 with a fixed short-range impurity 
density $n_d=2\times 10^{10}\ $cm$^{-2}$ and a tunable $n_d$ (shown in the inset),
respectively.
We also show the dependence of $\tau_s$ on $n_d$ (with the scale on the top of 
the frame) at $T=50\ $K with other conditions the same as the case of 
V$_{\rm CNP}+20\ $V. The short-range potential strength $V_0=6\times 10^{-17}\ $meV\,m$^2$.}
\label{fig5}
\end{figure}

\subsection{Comparison with experiments}
As mentioned in the introduction, experiments on spin relaxation have been carried out in BLG
on SiO$_2$ substrate\cite{yang,han107,avsar} or in freely suspended BLG\cite{neumann}
very recently by different groups. The in-plane SRTs in these experiments are 
of the order of $0.01$-$1\ $ns, which
are comparable to our theoretical results at high temperature (several hundred picoseconds).
This motivates us to carry out a detailed comparison with the existing 
experiments. In these experiments, the gate voltage, temperature and electron
density\cite{ed} (tuned by the gate voltage) are given explicitly. However, some parameters
such as the out-of-plane electric field and the impurity density are not available 
in the experiments. In our calculation, the out-of-plane electric 
field is taken to be the ratio of
the gate voltage to the thickness of the substrate approximately.\cite{hernando74,pzhang84}
The impurity density is obtained by fitting to the mobility or the spin diffusion coefficient
given in the experiments.\cite{sdc} 
In addition, we assume the initial spin polarization along the $x$-axis
without loss of generality since the in-plane spin
relaxations are identical as pointed out previously.

We first compare the temperature dependence of the SRTs obtained from the experiments and calculation from
our model. In these experiments, the mobilities are
reported of the order of $10^2$-$10^4\ $cm$^2$/(V\,s). 
The SRTs from different groups are comparable as shown in Fig.~\ref{fig5}.
However, the SRT from the experiment of Han and Kawakami\cite{han107} decreases with increasing temperature whereas the ones
from the experiments of Yang {\it et al.}\cite{yang} and Avsar {\it et al.}\cite{avsar} show a marginal temperature dependence.
In addition, Han and Kawakami\cite{han107} reported the longest SRT among these experiments, indicating that
their experimental data may contain less extrinsic effects. Motivated by this, we show
our comparison with the experiment of Han and Kawakami\cite{han107} in Fig.~\ref{fig5}(a).
The experimental data with the gate voltage V$_{\rm CNP}+20\ (60)\ $V corresponding to
the electron density $N_e=1.47\ (4.4)\times 10^{12}\ $cm$^{-2}$, is
labeled as $+\ (\blacktriangle)$. By fitting to the spin diffusion coefficient given in the experiment,
we obtain the long-range impurity density $N_i$ shown in the inset as the curve with $\bullet$. 
In contrast to the long-range impurity scattering, the 
contribution of other 
scatterings to the spin diffusion coefficient (mobility) is marginal.
Then we calculate the SRT corresponding to V$_{\rm CNP}+20\ (60)\ $V shown as the
red solid (orange chain) curve. 
A crossover is observed in the temperature dependence of the SRT, 
similar to the one shown in Sec.~IIIA. It is seen that the SRT 
from our calculation is comparable to the experimental data at high temperature. 
This suggests that the intervalley spin relaxation channel induced by the intervalley
electron-phonon scattering and the Zeeman-like term in
the two valleys addressed in Sec.~IIIA, plays an important role at high temperature. 
Nevertheless, the SRTs are still larger than the experimental data with 
an observable difference at high temperature.
Additionally, at low temperature, our results are orders of magnitude larger than the experimental ones since
the intervalley electron-phonon scattering is marginal at low temperature
and the SRT is determined by the long-range electron-impurity scattering
in our calculation. Due to the existence of all these discrepancies, one has 
to take into account extrinsic effects such as 
adatoms,\cite{abdelouahed,varykhalov,neto,ertler,dugaev,pzhang84,pzhang14}
curvature,\cite{morozov,jeong,hernando74,pzhang112}
substrate effects\cite{ryu,dedkov,ertler} and also 
contacts.\cite{popinciuc,han324,maassen,gallo} In the above 
experiments on BLG, the spin polarized electrons 
are all injected from ferromagnetic 
contacts.\cite{yang,han107,avsar,neumann} As reported, the ferromagnetic electrodes generate stray fields, 
which cause spin precession around the contact.\cite{gallo} 
As a result, the stray fields may have a strong influence on the observed experimental results. 
Additionally, electron scattering by the tunneling contacts may also be responsible for 
the available experimental data.\cite{popinciuc,han324,maassen}

In addition to the above extrinsic effects, the short-range 
scatterers may also be an extrinsic 
source.\cite{sarma81,adam82,sarma} Here, we discuss the contribution of short-range 
scatterers\cite{sarma} in detail. The scattering 
term induced by the short-range scatterers is given by
\begin{eqnarray}
\partial_t\hat{\rho}_{\mu{\bf k}}|_{\rm SR}&=&-\pi n_dV_0^2\sum_{\mu^{\prime}{\bf k}^{\prime}}
|{\psi_{\bf k}^{\mu}}^{\dagger}\psi_{{\bf k}^{\prime}}^{\mu^{\prime}}|^2\delta (\epsilon_{\mu^{\prime}{\bf k}^{\prime}}-\epsilon_{\mu {\bf k}})\nonumber\\
&&\mbox{}\hspace{-0.3cm}\times (\hat{\rho}_{\mu {\bf k}}-\hat{\rho}_{\mu^{\prime}{\bf k}^{\prime}})+{\rm H.c.},\label{eq16}
\end{eqnarray}
with $n_d$ and $V_0$ denoting the short-range impurity density and the constant short-range potential strength,
respectively.\cite{sarma} This scattering term contributes to both the intra- and inter-valley scatterings.
The intervalley scattering opens an intervalley spin relaxation channel 
together with the Zeeman-like term in the two valleys [see Eq.~(\ref{eq3})], 
similar to the intervalley electron-phonon scattering. However, in contrast to the intervalley
electron-phonon scattering, the intervalley scattering induced by the short-range scatterers is {\em insensitive} to
the temperature and may play an important role in the in-plane 
spin relaxation especially at low temperature.
Here, we choose a fixed short-range impurity density $n_d=2\times 10^{10}\ $cm$^{-2}$ and potential strength
$V_0=6\times 10^{-17}\ $meV\,m$^2$. 
It is noted that although with this additional short-range scattering included,
the mobility is still dominated by the long-range 
electron-impurity scattering since the contribution of the long-range electron-impurity
scattering is about four orders of magnitude larger than that of the short-range one.
We calculate the temperature dependence of the SRT
corresponding to the case of V$_{\rm CNP}+20\ $V in the experiment of
Han and Kawakami\cite{han107} by including the additional short-range scattering.
The result is plotted as the grey dashed curve in Fig.~\ref{fig5}(b).
A crossover is also observed at high temperature due to the contribution of the intervalley electron-phonon scattering
as pointed out previously. By comparing our calculation with the experimental data labeled as $+$,
we find that our result becomes comparable at low temperature. Furthermore, by fitting to
the experimental data with tunable
short-range impurity density shown as the curve with $\circ$ in the inset,
our result labeled as the skyblue double-dotted chain curve 
agrees fairly well with the experimental data 
in the temperature regime lower than the crossover. 
As for the temperature higher than the crossover, our
result becomes even larger than the experimental one once the short-range scattering is included.
This is because that the intervalley electron-phonon scattering is in the strong scattering limit. With
the inclusion of the short-range scattering, the intervalley scattering becomes stronger and
the SRT becomes larger.
This discrepancy may be attributed to other extrinsic effects.

As mentioned above, the SRT can be tuned by the short-range impurity density. We show the
short-range impurity density dependence of the in-plane 
SRT at $T=50\ $K as the purple dotted curve in Fig.~\ref{fig5}(b),
with all other conditions remaining the same as the case of V$_{\rm CNP}+20\ $V.
A valley is also observed, similar to the case of the temperature dependence discussed above.
At low temperature, the intervalley electron-phonon scattering is negligible. The SRT is then determined by the
intervalley short-range scattering. The valley originates from the crossover between the
weak and strong intervalley short-range scattering limit by tuning the short-range impurity density.
At the crossover point, the SRT $\tau_{s}\approx 232\ $ps according to Eq.~(\ref{eq15}),
close to the value $252\ $ps shown in the figure. This crossover point gives the minimum of the SRT and
can be taken as a criterion for determining the possible contribution of the short-range scattering.
Specifically, the crossover point $\tau_{s}\approx 316$, $258$, $232$, $283$, and $249\ $ps corresponding to the case of the experimental data shown 
as $\blacktriangle$, $\times$, $\square$, $\blacktriangledown$ 
and $\blacksquare$, respectively. We find that both the experimental 
data of Han and Kawakami\cite{han107} and Avsar {\it et al.}\cite{avsar}
are larger than the crossover point whereas
the one of Yang {\it et al.}\cite{yang} is smaller. 
This suggests the experimental data of Han and Kawakami\cite{han107} and Avsar {\it et al.}\cite{avsar} can be explained by the short-range scatterers.
As for the experiment of Yang {\it et al.},\cite{yang} some other extrinsic effects have to be considered.

\begin{figure}[bth]
  \includegraphics[width=8.5cm]{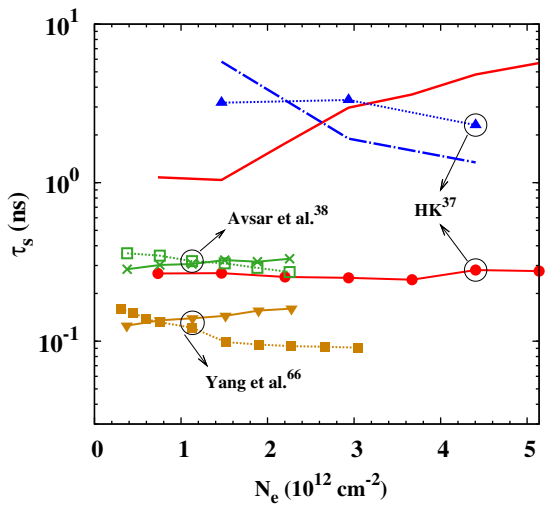}
\caption{(Color online) In-plane SRT $\tau_s$ as function of electron density.
The dotted curves with symbols correspond to the experimental data 
of Han and Kawakami (HK)\cite{han107}
at $T=20\ $K and Avsar {\it et al.}\cite{avsar} and Yang {\it et al.}\cite{yang} 
at $T=5\ $K. The solid curves with symbols correspond to the experimental data at room 
temperature of Han and Kawakami,\cite{han107} Avsar {\it et al.}\cite{avsar} and 
Yang {\it et al.}.\cite{yang} 
Red solid (blue chain) curve is from our calculation corresponding to the case
of room (low) temperature in the experiment of Han and Kawakami.\cite{han107} 
It is noted that the blue chain curve is calculated with 
the short-range scatterers with the short-range impurity density $n_d=10^{10}\ $cm$^{-2}$ 
and potential strength $V_0=6\times 10^{-17}\ $meV\,m$^2$.}
\label{fig6}
\end{figure}

Then we turn to compare the electron density dependence of 
the SRTs from the experiments and our calculation.
Among the experiments, Han and Kawakami\cite{han107} measured the longest SRT at low temperature 
(up to several nanoseconds) shown as the dotted curve with $\blacktriangle$ in Fig.~\ref{fig6}. 
This SRT is about one order of magnitude larger than the ones from other experimental cases. 
In addition, the SRTs from the experiments show different dependences on electron density. 
Specifically, at low temperature, both Avsar {\it et al.}\cite{avsar} and 
Yang {\it et al.}\cite{yang} show a decrease of the SRT with the 
increase of the electron density whereas a mild peak is observed by Han 
and Kawakami.\cite{han107} At room temperature, 
the SRTs from Avsar {\it et al.}\cite{avsar} 
and Yang {\it et al.}\cite{yang} increase with increasing electron density 
whereas Han and Kawakami\cite{han107} reported an insensitive density dependence 
of the SRT.

We first show our comparison with the experimental data 
at low temperature. As shown in Fig.~\ref{fig5}(a), our result is 
orders of magnitude larger than the experimental data of Han and Kawakami\cite{han107} 
at low temperature without short-range impurities. 
However, with the inclusion of the short-range scatterers, our result shows 
good agreement with the experimental data. Here, with the short-range scatterers 
explicitly included, we also show our comparison with the experimental data 
of Han and Kawakami.\cite{han107} With a fixed short-range impurity
density $n_d=10^{10}\ $cm$^{-2}$ and potential strength 
$V_0=6\times 10^{-17}\ $meV\,m$^2$,
the electron density dependence of the SRT from our calculation is 
shown as blue chain curve in Fig.~\ref{fig6}. We find that our 
result marginally agrees with the experimental data shown as the dotted curve with 
$\blacktriangle$. Better agreement can be obtained with tunable 
short-range impurity density. 
In addition, we also find that the experimental data of
Avsar {\it et al.}\cite{avsar} can be explained by the 
short-range scatterers. However, to account for the experimental data of 
Yang {\it et al.},\cite{yang} one has to include 
other extrinsic effects. This is because the SRT of Yang {\it et al.}\cite{yang} is 
smaller than the crossover point in the short-range impurity density dependence of the 
SRT ($230$-$320\ $ps), which is taken as a criterion for 
the possible contribution of the short-range scattering as mentioned previously.

We then address our comparison with the experimental data at room temperature. 
As shown in Fig.~\ref{fig5}(a), our result is larger than the 
experimental data of Han and Kawakami\cite{han107} with an observable difference 
at room temperature without short-range impurities. With the short-range scatterers 
included, the SRT from our calculation 
becomes even larger than the experimental data as mentioned previously, 
indicating that other extrinsic 
effects have to be taken into account. Here, 
we show our comparison with the experiment of Han and Kawakami\cite{han107} 
without extrinsic effects. The experimental data and our result are 
shown as the solid curve with $\bullet$ and the red solid curve in Fig.~\ref{fig6}, respectively. 
We find that the SRT in our calculation 
is about three times as large as the experimental one 
in low electron density regime and one order of magnitude 
larger than the experimental one in high density regime.

\section{SUMMARY}
In summary, we have investigated the electron spin relaxation due to the DP mechanism
in BLG with only the lowest conduction band being relevant. The SOC 
of the lowest conduction band is constructed from 
the symmetry group analysis with the 
parameters obtained by fitting to the numerical calculation according to 
the latest report by 
Konschuh {\it et al.}\cite{konschuh85} from first principles with 
both the intrinsic and extrinsic SOC terms in the pseudospin space included. 
We find that the magnitudes of both the out-of- 
and in-plane components of the 
 SOC decrease with increasing momentum at large momentum,
 indicating a suppression 
of the inhomogeneous broadening with the increase of the momentum. This is 
different from the case in both semiconductors and single-layer graphene. 
Additionally, the leading term of the
out-of-plane component of the SOC serves as a Zeeman-like term with opposite 
effective magnetic fields in the two valleys, 
which is similar to the case in rippled single-layer graphene. 
This Zeeman-like term, together with the 
intervalley electron-phonon and/or possible intervalley 
short-range scatterings, opens an intervalley spin relaxation channel,
which has not been reported in the literature in BLG. The intervalley
electron-phonon scattering is derived by using the tight-binding model. 
In addition to the intervalley electron-phonon and short-range
scatterings, we also include the long-range 
electron-impurity, 
electron-electron Coulomb 
and intravalley electron-phonon scatterings to calculate the SRT. We find that 
the in-plane SRT is strongly suppressed by the intervalley electron-phonon scattering 
at high temperature. In contrast to the intervalley electron-phonon 
scattering, the intervalley short-range scattering is insensitive to the temperature and 
plays an important role in the in-plane spin relaxation especially at low temperature. 

In the absence of short-range scatterers, a marked nonmonotonic temperature 
dependence of the in-plane SRT is predicted with a minimum SRT 
down to several hundred picoseconds. This nonmonotonic behavior originates 
from the crossover between the weak and strong intervalley electron-phonon scattering. 
Moreover, we predict a peak in the electron density dependence of the in-plane 
SRT at low temperature, which is very different from the one in semiconductors. 
At high temperature, the in-plane SRT increases 
monotonically with increasing density. We also find that the 
in-plane SRT decreases rapidly with the increase of the 
initial spin polarization at low temperature. 
This is very different from the previous studies 
in both semiconductors and single-layer graphene where the SRT increases significantly 
with increasing initial spin polarization. The physics is understood 
that the spin relaxation time is determined by the intervalley electron-phonon 
scattering, which transfers electrons between the two valleys. The transferred 
electrons experience opposite effective magnetic fields in the two valleys, which 
are not affected by the Coulomb HF term. 
In addition, a strong anisotropy between the out-of- and in-plane spin 
relaxations is also addressed at high temperature where the out-of-plane SRT is about 
two orders of magnitude larger than the in-plane one. As for low temperature, 
the out-of- and in-plane spin relaxation times are comparable.

We also show our comparison with the existing experiments 
of Han and Kawakami,\cite{han107} Avsar {\it et al.}\cite{avsar} and 
Yang {\it et al.}.\cite{yang} We find that without intervalley scattering, 
the SRT is orders of magnitude larger than the experimental data in the whole temperature regime. 
With the intervalley electron-phonon scattering explicitly included, the 
SRT from our calculation becomes comparable to the experimental data at high temperature 
but still orders of magnitude larger than the experimental data 
at low temperature. In addition, a crossover in the temperature 
dependence of the in-plane SRT is shown, which also results from the crossover 
between the weak and strong intervalley electron-phonon scattering. With 
the inclusion of the short-range scatterers, our result agrees fairly well 
with the experimental data in the temperature regime 
lower than the crossover. As for the temperature higher 
than the crossover, other extrinsic effects have to be included. Moreover, 
a crossover point is also shown in the short-range impurity density dependence of the 
in-plane SRT, which gives a minimum for determining the possible contribution 
of the short-range scattering. The experimental SRT larger than the minimum can 
be explained by the possible short-range scatterers whereas other extrinsic effects have 
to be considered for experimental data smaller than the minimum.

\begin{acknowledgments}
This work was supported by the National Basic Research Program of
China under Grant No.\ 2012CB922002 and the Strategic 
Priority Research Program of the
Chinese Academy of Sciences under Grant No. XDB01000000. One of the
 authors (L.W.) would like
to thank B. Y. Sun for double checking the derivation of the 
electron-phonon scattering matrix
elements.
\end{acknowledgments}

\begin{figure}[thp]
\includegraphics[width=6.8cm]{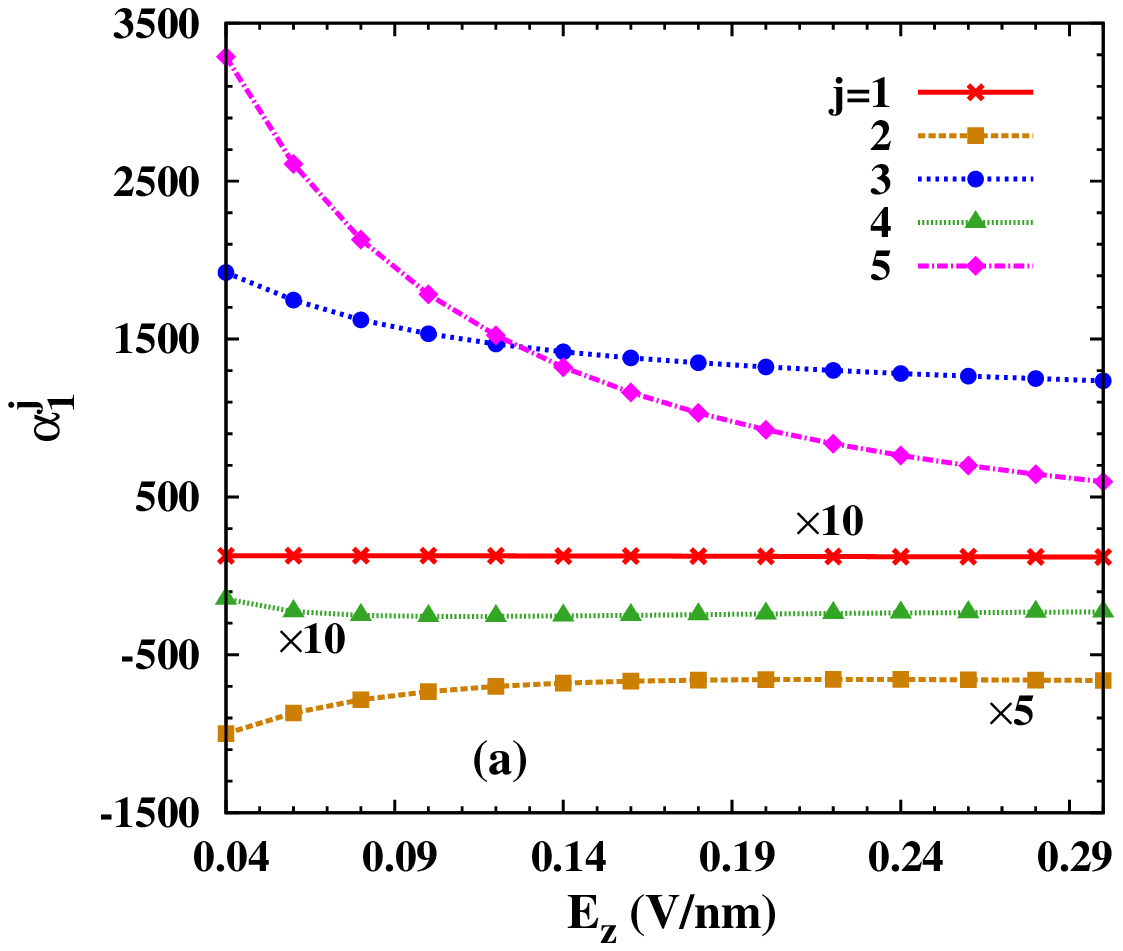}
\includegraphics[width=6.8cm]{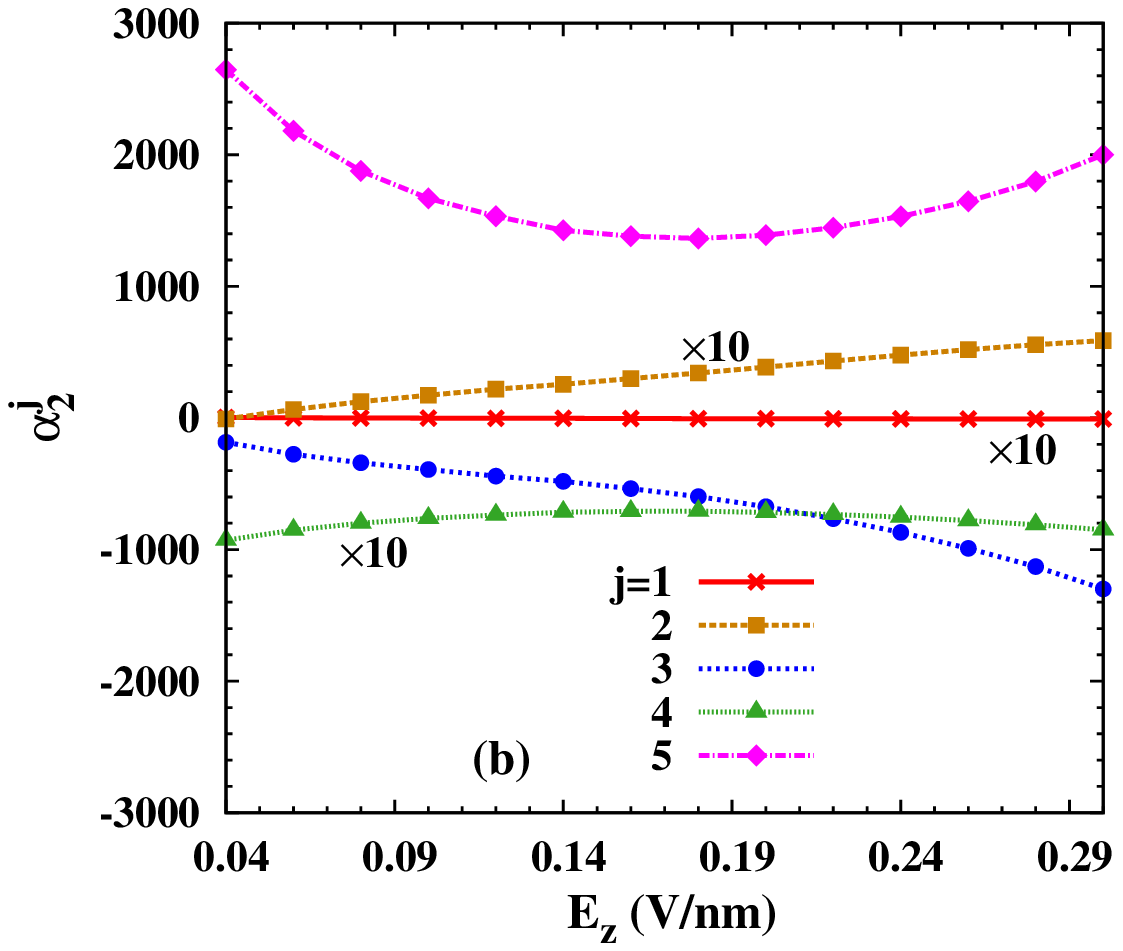}
\includegraphics[width=6.8cm]{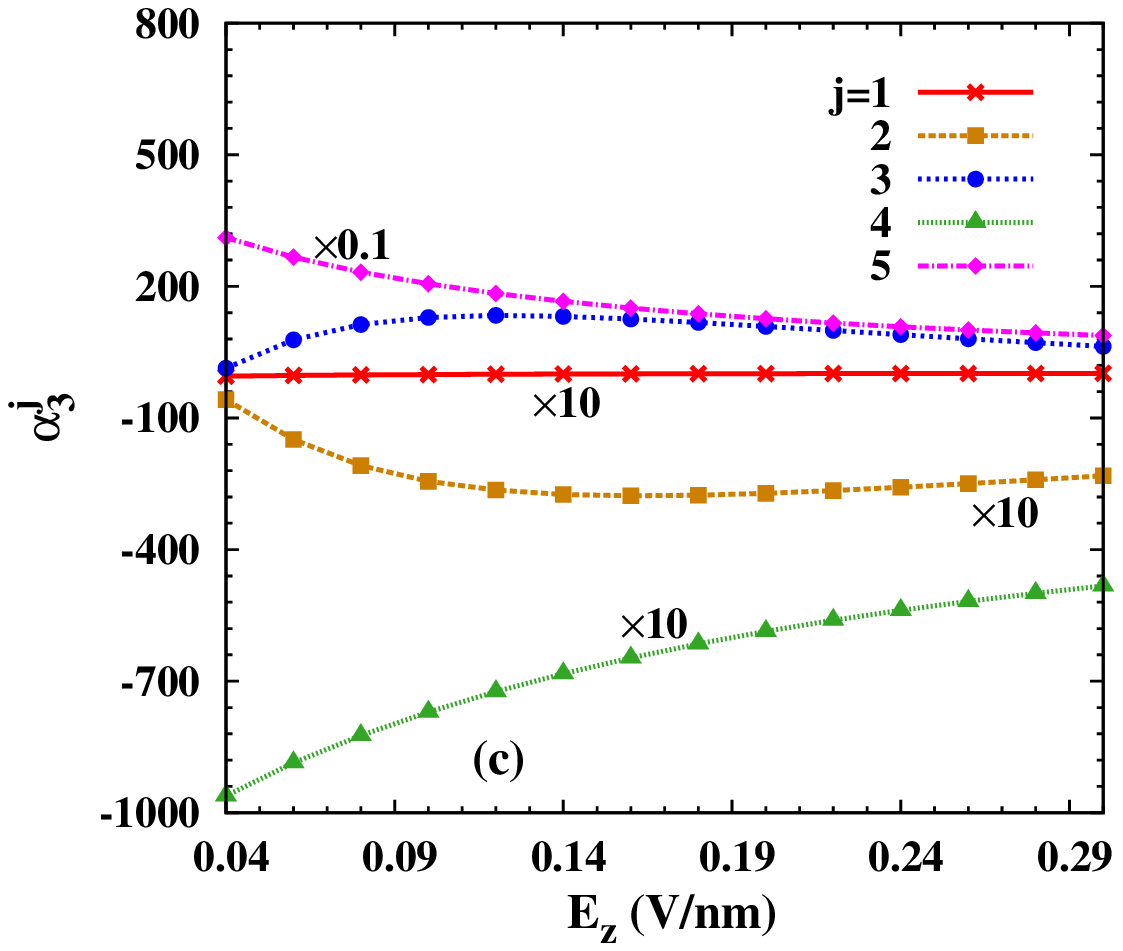}
\caption{(Color online) (a) $\alpha_1^j$, (b) $\alpha_2^j$ and (c) $\alpha_3^j$ 
as function of the out-of-plane electric field $E_z$ with $j=1$-$5$. 
It is noted that $\alpha_i^1$, $\alpha_i^2$, 
$\alpha_i^3$, $\alpha_i^4$ and $\alpha_i^5$ ($i=1$-$3$) are dimensionless.}
\label{fig7}
\end{figure}

\begin{appendix}
\section{SOC OF THE LOWEST CONDUCTION BAND}\label{appA}
The SOC term of the lowest conduction band near the Dirac points can be described by a $2\times 2$ matrix
\begin{eqnarray}
H_{{\rm eff}\mu}^{\rm SO}=\left(\begin{array}{cc}
H_{\mu 11}^{\rm SO} & H_{\mu 12}^{\rm SO} \\
{H_{\mu 12}^{\rm SO}}^* & H_{\mu 22}^{\rm SO} \\
\end{array}\right)\label{A1}
\end{eqnarray}
in the basis $\psi^{\mu}_{{\bf k}s}=\psi^{\mu}_{{\bf k}}\otimes |s\rangle$ where
$s=\{\uparrow,\downarrow\}$ and $\psi^{\mu}_{{\bf k}}=\sum_ic_i^{\mu}({\bf k})\Psi_i^{\mu}({\bf k})$
($i={\rm A_1,\ B_1,\ A_2,\ B_2}$). The coefficients $c_i^{\mu}({\bf k})$ can be obtained
by exactly diagonalizing the $4\times 4$ effective Hamiltonian given in Eq.~(\ref{eq5}). Here,
\begin{eqnarray}
H_{\mu 11}^{\rm SO}&=&\sum_{i,j}{c_i^{\mu}}^*({\bf k})c_j^{\mu}({\bf k})\langle\Psi_{i,\uparrow}^{\mu}|H_{\rm SO}|
\Psi_{j,\uparrow}^{\mu}\rangle,\label{A2}\\
H_{\mu 22}^{\rm SO}&=&\sum_{i,j}{c_i^{\mu}}^*({\bf k})c_j^{\mu}({\bf k})\langle\Psi_{i,\downarrow}^{\mu}|H_{\rm SO}|
\Psi_{j,\downarrow}^{\mu}\rangle,\label{A3}\\
H_{\mu 12}^{\rm SO}&=&\sum_{i,j}{c_i^{\mu}}^*({\bf k})c_j^{\mu}({\bf k})\langle\Psi_{i,\uparrow}^{\mu}|H_{\rm SO}|
\Psi_{j,\downarrow}^{\mu}\rangle,\label{A4}
\end{eqnarray}
with $\langle\Psi_{i,s_1}^{\mu}|H_{\rm SO}|\Psi_{j,s_2}^{\mu}\rangle$ 
($s_{1,2}=\{\uparrow,\downarrow\}$) 
being the spin-orbit matrix elements near the Dirac points 
in the pseudospin space given by Konschuh {\it et al.}.\cite{konschuh85} 
Specifically, 
\begin{eqnarray}
H_{\mu 11}^{\rm SO}&=&\mu\lambda_{{\rm I}1}(|c_3^{\mu}|^2-|c_2^{\mu}|^2)+\mu\lambda_{{\rm I}2}(|c_1^{\mu}|^2-|c_4^{\mu}|^2)\label{A5}\\
H_{\mu 22}^{\rm SO}&=&-H_{\mu 11}^{\rm SO}\label{A6}\\
H_{\mu 12}^{\rm SO}&=&\frac{\mu+1}{2}[{c_1^{\mu}}^*c_3^{\mu}(i\lambda_4)+{c_2^{\mu}}^*c_1^{\mu}(-i\lambda_0)+{c_2^{\mu}}^*c_4^{\mu}(-i\lambda_4^{\prime})\nonumber\\
&&\mbox{}\hspace{-0.4cm}+{c_3^{\mu}}^*c_2^{\mu}(-i\lambda_3)+{c_4^{\mu}}^*c_3^{\mu}(i\lambda_0^{\prime})]+\frac{\mu-1}{2}[{c_3^{\mu}}^*c_1^{\mu}(i\lambda_4)
\nonumber\\
&&\mbox{}\hspace{-0.4cm}+{c_1^{\mu}}^*c_2^{\mu}(-i\lambda_0)+{c_4^{\mu}}^*c_2^{\mu}(-i\lambda_4^{\prime})+{c_2^{\mu}}^*c_3^{\mu}(-i\lambda_3)\nonumber\\
&&\mbox{}\hspace{-0.4cm}+{c_3^{\mu}}^*c_4^{\mu}(i\lambda_0^{\prime})].\label{A7}
\end{eqnarray}
$\lambda_{{\rm I}1}$ and $\lambda_{{\rm I}2}$ are the strengths of single-layer-like intrinsic SOC;
$\lambda_0$ and $\lambda_0^{\prime}$ represent single-layer-like extrinsic spin-orbit strengths; $\lambda_3$, $\lambda_4$ and
$\lambda_4^{\prime}$ stand for interlayer spin-orbit parameters.\cite{konschuh85}
Then the effective magnetic field of the SOC ${\bf\Omega}^{\mu}({\bf k})$ 
in Eq.~(\ref{eq6}) is given by $\Omega_x^{\mu}({\bf k})=2{\rm Re}H_{\mu 12}^{\rm SO}$, 
$\Omega_y^{\mu}({\bf k})=-2{\rm Im}H_{\mu 12}^{\rm SO}$
and $\Omega_z^{\mu}({\bf k})=2H_{\mu 11}^{\rm SO}$. 
It is noted that the contribution of other energy bands to the
SOC of the lowest conduction band is marginal in our calculation.

\begin{figure}[bth]
\includegraphics[width=7.5cm]{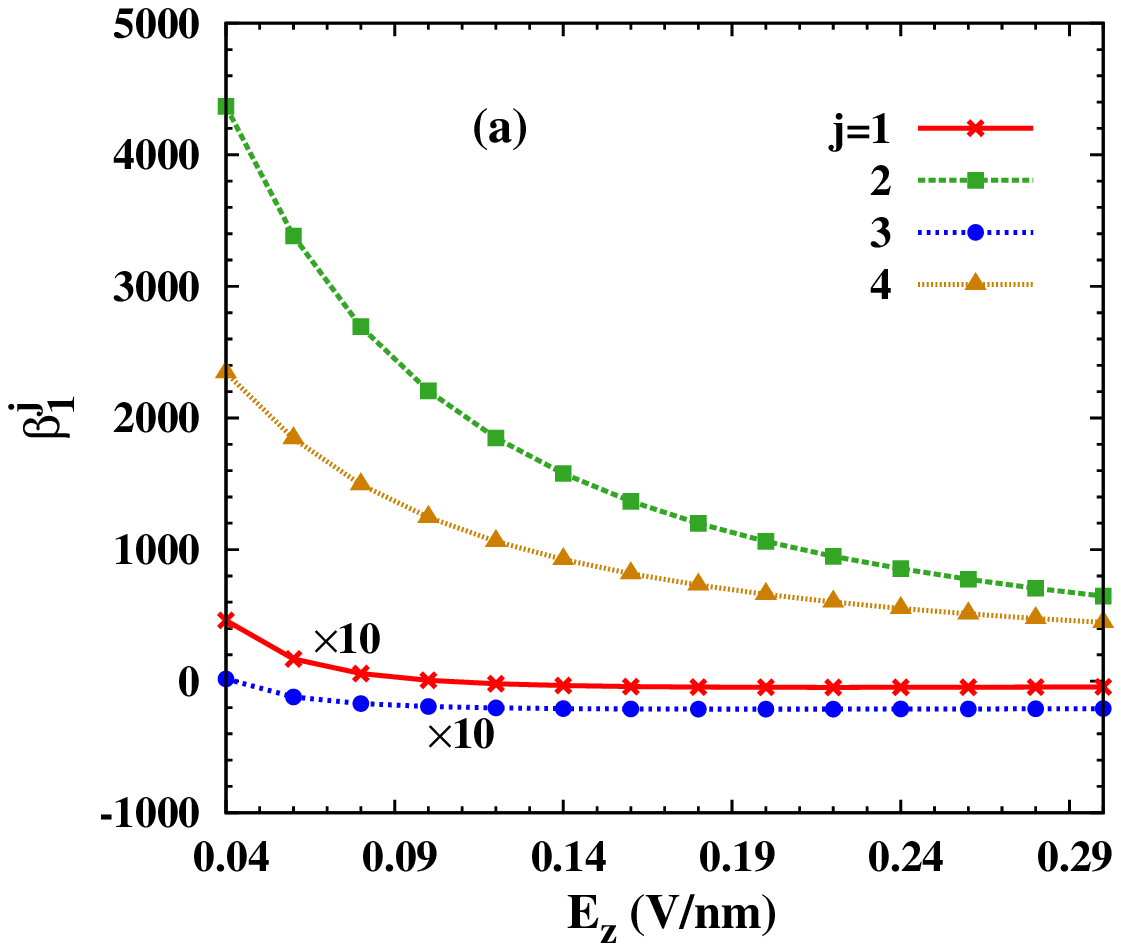}
\includegraphics[width=7.5cm]{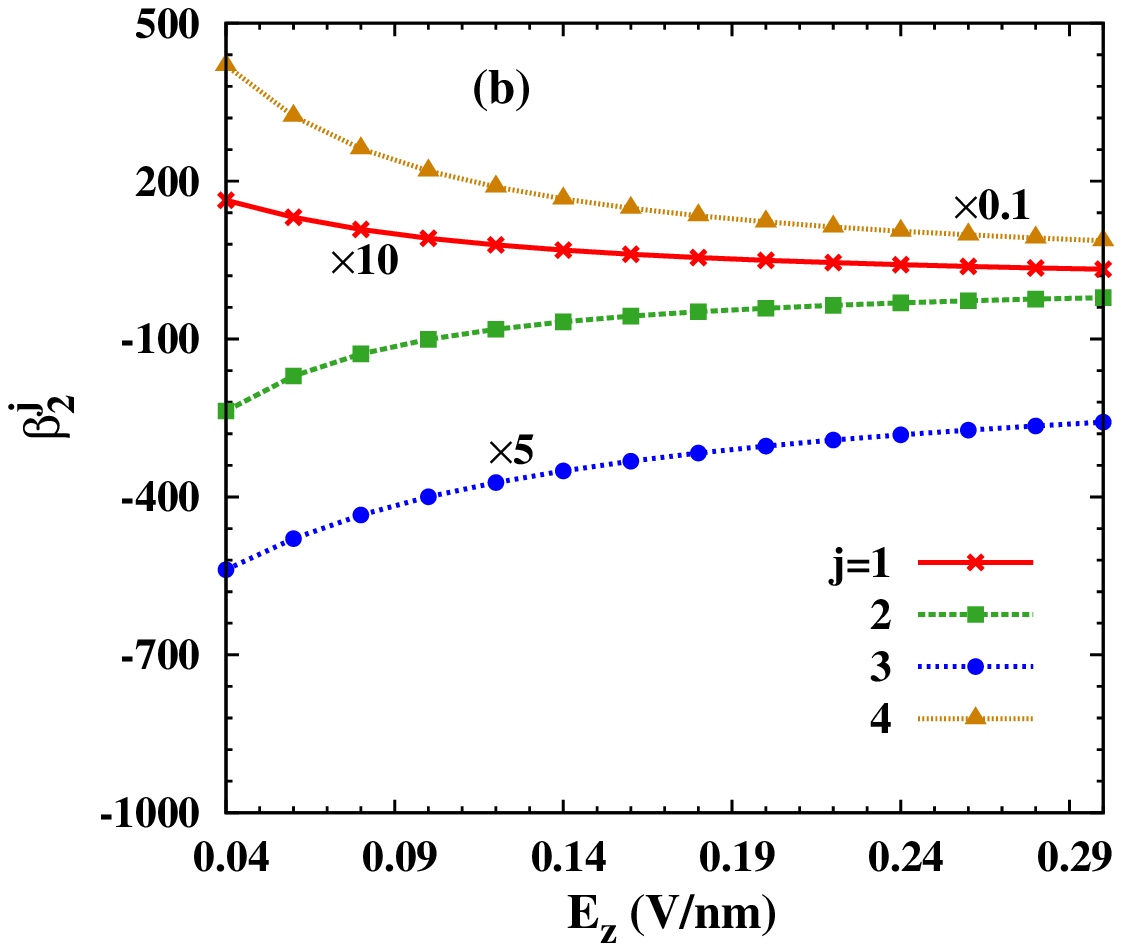}
\caption{(Color online) (a) $\beta_1^j$ and (b) $\beta_2^j$ as function of 
the out-of-plane electric field $E_z$ with $j=1$-$4$. 
It is noted that $\beta_i^1$, $\beta_i^2$, $\beta_i^3$ and $\beta_i^4$ 
($i=1,2$) are dimensionless.}
\label{fig8}
\end{figure}

As reported by Konschuh {\it et al.},\cite{konschuh85} the small group of the Dirac points is $C_3$
in the presence of an out-of-plane electric field. Following from
the symmetry group $C_3$,\cite{koster} we derive an analytical form of the
SOC near the Dirac points [see Eqs.~(\ref{eq1}-\ref{eq3})]. In these equations,
the coefficients $\alpha_i(k)$ ($i=1$-$3$) and $\beta_i(k)$ 
($i=1$-$2$) are obtained  by 
fitting to the numerical calculation from Eqs.~(\ref{A5}) and (\ref{A7}) 
using the Pad\'{e} approximation.\cite{pade} Specifically,
the coefficients $\alpha_i(k)$ ($i=1$-$3$) in Eqs.~(\ref{eq1}-\ref{eq2}) read
\begin{eqnarray}
\alpha_i(k)&=&\lambda_{{\rm I}1}ak\frac{\alpha_i^1+\alpha_i^2ak+\alpha_i^3a^2k^2}{1+\alpha_i^4ak+\alpha_i^5a^2k^2}.\label{A8}
\end{eqnarray}
The coefficients $\beta_i(k)$ ($i=1$-$2$) in Eq.~(\ref{eq3}) are given by
\begin{eqnarray}
\beta_1(k)&=&\beta_1^0+\lambda_{{\rm I}1}ak\frac{\beta_1^1+\beta_1^2ak}{1+\beta_1^3ak+\beta_1^4a^2k^2},\label{A9}\\
\beta_2(k)&=&\lambda_{{\rm I}1}ak\frac{\beta_2^1+\beta_2^2ak}{1+\beta_2^3ak+\beta_2^4a^2k^2}.\label{A10}
\end{eqnarray}
Here, the coefficient $\beta_1^0=-24\ \mu$eV, which is independent of the applied electric field.
The electric field dependences of $\alpha_i^j$ ($i=1$-$3$, $j=1$-$5$) and $\beta_{i}^{j}$ ($i=1$-$2$,
$j=1$-$4$) are plotted in Figs.~\ref{fig7} and \ref{fig8}, respectively. 
It is noted that the Pad\'{e} approximation\cite{pade} can 
give a precise description of 
$\alpha_i(k)$ ($i=1$-$3$) and $\beta_i(k)$ ($i=1$-$2$) at large 
momenta near the Dirac points, which can be populated and 
play an important role in spin relaxation according to the experimental 
conditions (heavily doped).\cite{yang,han107,avsar,neumann} 
With only a simple linear $k$-order 
approximation, $\alpha_i(k)$ and $\beta_i(k)$ agree with the 
numerical results only at very small momentum near the Dirac points.

\begin{figure}[bth]
\includegraphics[width=7.5cm]{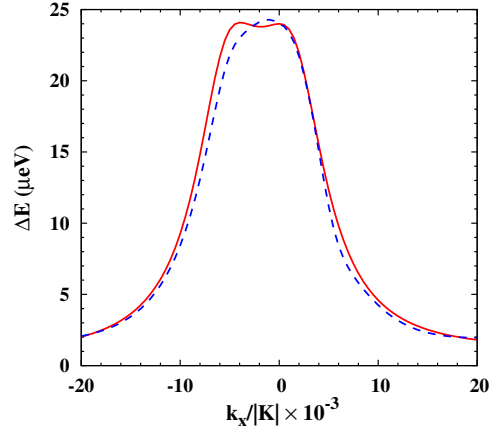}
\caption{(Color online) Spin splitting of the lowest conduction 
band $\Delta E$ near the K point as function of
the momentum $k_x$ under a typical electric field $E_z=0.14\ $V/nm. 
Solid (dashed) curve represents
the numerical (analytical) result. $|{\rm K}|=4\pi/(3a)$.}
\label{fig9}
\end{figure}

We also show in Fig.~\ref{fig9} the spin splitting of the lowest conduction band calculated
with analytical form of the SOC (blue dashed curve) and with the explicit numerical one (red solid curve). We
find that the analytical result agrees fairly well with the numerical one at large momentum. In addition,
the spin splitting decreases with increasing momentum
 at large momentum, which is very different from the
case in both semiconductors\cite{wureview} and single-layer graphene.\cite{yzhou}

\section{SCATTERING MATRIX ELEMENTS}\label{appB}
The electron-electron Coulomb scattering matrix element is 
given by $|V_{{\bf k},{\bf {k-q}}}^{\mu}|^2$.
The screened Coulomb potential $V_{{\bf k},{\bf {k-q}}}^{\mu}=
V_{\bf q}^{(0)}/\varepsilon({\bf q},
\epsilon_{\mu{\bf k}}-\epsilon_{\mu{\bf k-q}})$ with
$\varepsilon({\bf q},\epsilon_{\mu{\bf k}}-\epsilon_{\mu{\bf k-q}})=
1-V_{\bf q}^{(0)}\Pi({\bf q},\epsilon_{\mu{\bf k}}-\epsilon_{\mu{\bf k-q}})$
being the screening under the random phase approximation.\cite{mahan} 
$V_{\bf q}^{(0)}=2\pi v_{\rm F}r_s/q$ is
the two-dimensional bare Coulomb potential with $v_{\rm F}$ 
and $r_s$ being the Fermi velocity in
single-layer graphene\cite{divincenzo} and the dimensionless Wigner-Seitz radius,\cite{adam1,hwang,adam2} respectively.
$\Pi({\bf q},\omega)$ is given by\cite{yzhou,wunsch,xfwang,hwang75}
\begin{eqnarray}
\Pi({\bf q},\omega)=\sum_{\mu{\bf k}\nu\nu^{\prime} s}|T_{{\bf k}{\bf k+q}}^{\mu\nu\nu^{\prime}}|^2\frac{f_{{\bf k}s}^{\mu\nu}-f_{{\bf k+q}s}^{\mu\nu^{\prime}}}
{\epsilon_{\mu\nu{\bf k}}-\epsilon_{\mu\nu^{\prime}{\bf k+q}}+\omega+i0^+},\ \
\end{eqnarray}
where $T_{{\bf k}{\bf k+q}}^{\mu\nu\nu^{\prime}}={\psi_{\bf k}^{\mu\nu}}^{\dagger}\psi_{\bf k+q}^{\mu\nu^{\prime}}$. The long-range electron-impurity
scattering matrix element $|U_{{\bf k},{\bf k-q}}^{\mu}|^2=Z_i^2|V_{{\bf k},{\bf {k-q}}}^{\mu}|^2e^{-2qd}$, in which $Z_i$
and $d$ stand for the impurity charge number and effective 
distance of the impurity layer to the BLG
sheet.\cite{ertler,fratini,adam1,hwang,adam2} It is noted that 
when we calculate the electron-electron
Coulomb and long-range 
electron-impurity scatterings, we take the distance between the graphene layers
to be zero approximately.\cite{gamayun}
It is further noted that the Coulomb potential in the HF term is in the static screening limit, i.e.,
$V_{{\bf k},{\bf {k-q}}}^{\mu}=V_{\bf q}^{(0)}/\varepsilon({\bf q},0)$.

Then we turn to the electron-phonon scattering matrix elements.
For the intravalley electron-AC-phonon scattering, the scattering matrix element
$|M^{\rm AC}_{{\mu{\bf k}},{\mu^{\prime}{\bf k}^{\prime}}}|^2=D_{\rm AC}^2\hbar 
q/(2\rho v_{\rm ph})
I_{{\bf k}{\bf k}^{\prime}}^{\mu}\delta_{\mu\mu^{\prime}}$ where $q=|{\bf k}
-{\bf k}^{\prime}|$;
$D_{\rm AC}$ and $v_{\rm ph}$ represent the deformation potential\cite{borysenko} and
acoustic phonon velocity,\cite{hwang2,jhchen} respectively.

For the intravalley electron-RI-phonon scattering,
$|M^{\rm RI}_{{\mu{\bf k}},{\mu^{\prime}{\bf k}^{\prime}}}|^2=g\frac{\sqrt{3}\hbar^2v_{\rm F}^2}{a}\frac{e^{-2qd}}{q+q_s}
I_{{\bf k}{\bf k}^{\prime}}^{\mu}\delta_{\mu\mu^{\prime}}$ with $q_s=4r_sk_{\rm F}$ being the Thomas-Fermi
screening length.\cite{wunsch} For SiO$_2$ substrate, the energy spectra 
of the two remote phonon modes
are denoted as $\omega_1^{\rm RI}$ and $\omega_2^{\rm RI}$, respectively; $g_1$ and $g_2$ are
the corresponding dimensionless coupling parameters.\cite{fratini}

The matrix elements for the intravalley electron-OP-phonon scattering ($|M^{\rm OP}_{{\mu{\bf k}},{\mu^{\prime}{\bf k}^{\prime}}}|^2$)
are derived using the tight-binding model
according to the arXiv version of Ref.~\onlinecite{viljas}.
The intravalley electron-OP-phonon scattering includes both the electron in-plane
($|M^{\rm LT}_{{\mu{\bf k}},{\mu^{\prime}{\bf k}^{\prime}}}|^2$) and out-of-plane
($|M^{\rm ZO}_{{\mu{\bf k}},{\mu^{\prime}{\bf k}^{\prime}}}|^2$) OP phonon scatterings. The
electron in-plane OP phonon scattering matrix element is given by
\begin{eqnarray}
|M^{\rm LT}_{{\mu{\bf k}},{\mu^{\prime}{\bf k}^{\prime}}}|^2&=&\frac{9\hbar^2}
{2\rho\Omega_{\rm LT}}\delta_{\mu\mu^{\prime}}
\Big\{|{\psi^{\mu}_{{\bf k}}}^{\dagger}[\gamma_0^{\prime}
(\sin\theta_{\bf q}\sigma^{23}_{\rm D}+\cos\theta_{\bf q}\sigma^{13}_{\rm D})\nonumber\\
&&\mbox{}\hspace{-1.9cm}-a\gamma_3^{\prime}(\sin\theta_{\bf q}\gamma^1_{\rm D}\gamma^5_{\rm D}-i\sin\theta_{\bf q}\gamma^2_{\rm D}+
i\cos\theta_{\bf q}\gamma^1_{\rm D}+\cos\theta_{\bf q}\nonumber\\
&&\mbox{}\hspace{-1.9cm}\times\gamma^2_{\rm D}\gamma^5_{\rm D})/(2\sqrt{3}l_3)]\psi^{\mu}_{{\bf k}^{\prime}}|^2+|{\psi^{\mu}_{{\bf k}}}^{\dagger}
[i\gamma_0^{\prime}(\sin\theta_{\bf q}\sigma^{01}_{\rm D}-\cos\theta_{\bf q}\nonumber\\
&&\mbox{}\hspace{-1.9cm}\times\sigma^{02}_{\rm D})+a\gamma_4^{\prime}(i\cos\theta_{\bf q}\gamma^3_{\rm D}-
\sin\theta_{\bf q}\gamma^3_{\rm D}\gamma^5_{\rm D})/(\sqrt{3}l_4)]\psi^{\mu}_{{\bf k}^{\prime}}|^2\nonumber\\
&&\mbox{}\hspace{-1.9cm}+|{\psi^{\mu}_{{\bf k}}}^{\dagger}[\gamma_0^{\prime}(\cos\theta_{\bf q}\sigma^{23}_{\rm D}
-\sin\theta_{\bf q}\sigma^{13}_{\rm D})-a\gamma_3^{\prime}(\cos\theta_{\bf q}\gamma^1_{\rm D}\gamma^5_{\rm D}\nonumber\\
&&\mbox{}\hspace{-1.9cm}-i\cos\theta_{\bf q}\gamma^2_{\rm D}-i\sin\theta_{\bf q}\gamma^1_{\rm D}-\sin\theta_{\bf q}\gamma^2_{\rm D}
\gamma^5_{\rm D})/(2\sqrt{3}l_3)]\psi^{\mu}_{{\bf k}^{\prime}}|^2\nonumber\\
&&\mbox{}\hspace{-1.9cm}+|{\psi^{\mu}_{{\bf k}}}^{\dagger}[i\gamma_0^{\prime}(\cos\theta_{\bf q}\sigma^{01}_{\rm D}
+\sin\theta_{\bf q}\sigma^{02}_{\rm D})-a\gamma_4^{\prime}\gamma^3_{\rm D}(i\sin\theta_{\bf q}\nonumber\\
&&\mbox{}\hspace{-1.9cm}+\cos\theta_{\bf q}\gamma^5_{\rm D})/(\sqrt{3}l_4)]\psi^{\mu}_{{\bf k}^{\prime}}|^2\Big\},
\end{eqnarray}
where $\Omega_{\rm LT}$ is the energy of the in-plane OP phonon modes;\cite{piscanec,lazzeri}
$l_3$ is the bond length corresponding to the interlayer hopping
$\gamma_3$; and $\sigma^{02}_{\rm D}$ and $\sigma^{13}_{\rm D}$ are
$4\times 4$ Dirac matrices given in Appendix~\ref{appC}.\cite{peskin} Here,
$\theta_{\bf q}$ is the polar angle of the momentum ${\bf q}$.
The electron out-of-plane OP phonon scattering matrix element reads
\begin{eqnarray}
|M^{\rm ZO}_{{\mu{\bf k}},{\mu^{\prime}{\bf k}^{\prime}}}|^2&=&\frac{{\hbar^2\gamma_1^{\prime}}^2}{2\rho\Omega_{\rm ZO}}\delta_{\mu\mu^{\prime}}
|{\psi^{\mu}_{{\bf k}}}^{\dagger}(\gamma^1_{\rm D}\gamma^5_{\rm D}+i\gamma^2_{\rm D})\psi^{\mu}_{{\bf k}^{\prime}}|^2,\nonumber\\
\end{eqnarray}
with $\Omega_{\rm ZO}$ being the energy of the out-of-plane OP phonon mode.\cite{borysenko}

\section{RELEVANT DIRAC MATRICES}\label{appC}

The relevant Dirac matrices\cite{peskin} used in the electron-phonon scattering are
\begin{equation}
\gamma^0_{\rm D}=\left(\begin{array}{cc}
0 & I \\
I & 0  \\
\end{array}\right),
\hspace{1.1cm}\gamma^5_{\rm D}=\left(\begin{array}{cc}
-I & 0 \\
0 & I  \\
\end{array}\right),
\end{equation}
\begin{eqnarray}
&&\sigma^{01}_{\rm D}=-i\left(\begin{array}{cc}
\sigma_x & 0 \\
0 & -\sigma_x  \\
\end{array}\right),
\sigma^{13}_{\rm D}=-\left(\begin{array}{cc}
\sigma_y & 0 \\
0 & \sigma_y  \\
\end{array}\right),\\
&&\sigma^{02}_{\rm D}=-i\left(\begin{array}{cc}
\sigma_y & 0 \\
0 & -\sigma_y  \\
\end{array}\right),
\sigma^{23}_{\rm D}=\left(\begin{array}{cc}
\sigma_x & 0 \\
0 & \sigma_x  \\
\end{array}\right),
\end{eqnarray}
\begin{equation}
\gamma^i_{\rm D}=\left(\begin{array}{cc}
0 & \sigma_i \\
-\sigma_i & 0  \\
\end{array}\right)\ (i=1\mbox{-}3),
\end{equation}
with $I$ being $2\times 2$ unit matrix.
\end{appendix}


\begin{thebibliography}{0}
\bibitem{novoselov} K. S. Novoselov, A. K. Geim, S. V. Morozov, D. Jiang, Y. Zhang, S. V. Dubonos, I. V.
Grigorieva, and A. A. Firsov, Science {\bf 306}, 666 (2004).

\bibitem{geim6} A. K. Geim and K. S. Novoselov, Nature Mater. {\bf 6}, 183 (2007).

\bibitem{tombros448} N. Tombros, C. Jozsa, M. Popinciuc, H. T. Jonkman, and
B. J. van Wees, Nature (London) {\bf 448}, 571 (2007).

\bibitem{fwang} F. Wang, Y. Zhang, C. Tian, C. Girit, A. Zettl, M. Crommie, and Y. Ron Shen,
Science {\bf 320}, 206 (2008).

\bibitem{beenakker} C. W. J. Beenakker, Rev. Mod. Phys. {\bf 80}, 1337 (2008).

\bibitem{neto81} A. H. Castro Neto, F. Guinea, N. M. R. Peres, K. S. Novoselov, and A. K. Geim, Rev.
Mod. Phys. {\bf 81}, 109 (2009).

\bibitem{peres} N. M. R. Peres, Rev. Mod. Phys. {\bf 82}, 2673 (2010).

\bibitem{abergel} D. S. L. Abergel, V. Apalkov, J. Berashevich, K. Ziegler, and T. Chakraborty,
Adv. Phys. {\bf 59}, 261 (2010).


\bibitem{sarma} S. Das Sarma, S. Adam, E. H. Hwang, and E. Rossi, Rev. Mod. Phys. {\bf 83}, 407 (2011).



\bibitem{acik} M. Acik and Y. J. Chabal, Jpn. J. Appl. Phys.
{\bf 50}, 070101 (2011).

\bibitem{geim83} A. K. Geim, Rev. Mod. Phys. {\bf 83}, 851 (2011).

\bibitem{goerbig} M. O. Goerbig, Rev. Mod. Phys. {\bf 83}, 1193 (2011).

\bibitem{kotov} V. N. Kotov, B. Uchoa, V. M. Pereira, F. Guinea, and A. H. Castro Neto,
Rev. Mod. Phys. {\bf 84}, 1067 (2012).

\bibitem{cooper} D. R. Cooper, B. D'Anjou, N. Ghattamaneni, B. Harack, M. Hilke, A. Horth, N. Majlis, M. Massicotte,
L. Vands-burger, E. Whiteway, and V. Yu, ISRN Condensed Matter 
Physics {\bf 2012}, 501686 (2012).



\bibitem{kane} C. L. Kane and E. J. Mele, Phys. Rev. Lett. {\bf 95}, 226801 (2005).

\bibitem{hernando74} D. Huertas-Hernando, F. Guinea, and A. Brataas, Phys. Rev. B {\bf 74}, 155426 (2006).

\bibitem{min} H. Min, J. E. Hill, N. A. Sinitsyn, B. R. Sahu, L. Kleinman, and A. H. MacDonald, Phys.
Rev. B {\bf 74}, 165310 (2006).

\bibitem{yao} Y. Yao, F. Ye, X.-L. Qi, S.-C. Zhang, and Z. Fang, 
Phys. Rev. B {\bf 75}, 041401(R) (2007).

\bibitem{boettger} J. C. Boettger and S. B. Trickey, Phys. Rev. B {\bf 75}, 
121402(R) (2007).

\bibitem{trauzettel} B. Trauzettel, D. V. Bulaev, D. Loss, and G. Burkard, Nature Phys. {\bf 3}, 192 (2007).

\bibitem{fischer} J. Fischer, B. Trauzettel, and D. Loss, Phys. Rev. B {\bf 80}, 155401 (2009).

\bibitem{gmitra} M. Gmitra, S. Konschuh, C. Ertler, C. Ambrosch-Draxl, and J. Fabian, Phys. Rev. B {\bf 80},
235431 (2009).

\bibitem{hernando103} D. Huertas-Hernando, F. Guinea, and A. Brataas, Phys. Rev. Lett. {\bf 103}, 146801 (2009).


\bibitem{abdelouahed} S. Abdelouahed, A. Ernst, J. Henk, I. V. Maznichenko, and I. Mertig, Phys. Rev. B
{\bf 82}, 125424 (2010).

\bibitem{konschuh82} S. Konschuh, M. Gmitra, and J. Fabian, Phys. Rev. B {\bf 82}, 245412 (2010).


\bibitem{cho} S. Cho, Y.-F. Chen, and M. S. Fuhrer, Appl. Phys. Lett. {\bf 91}, 123105 (2007).

\bibitem{jozsa100} C. J\'{o}zsa, M. Popinciuc, N. Tombros, H. T. Jonkman, and B. J. van Wees,
Phys. Rev. Lett. {\bf 100}, 236603 (2008).



\bibitem{tombros101} N. Tombros, S. Tanabe, A. Veligura, 
C. J\'{o}zsa, M. Popinciuc, H. T. Jonkman, and
B. J. van Wees, Phys. Rev. Lett. {\bf 101}, 046601 (2008).

\bibitem{jozsa80} C. J\'{o}zsa, T. Maassen, M. Popinciuc, P. J. Zomer, 
A. Veligura, H. T. Jonkman, and B. J. van Wees, Phys. Rev. B {\bf 80}, 241403(R)
 (2009).

\bibitem{popinciuc} M. Popinciuc, C. J\'{o}zsa, P. J. Zomer, N. Tombros, A. Veligura, H. T. Jonkman, and
B. J. van Wees, Phys. Rev. B {\bf 80}, 214427 (2009).

\bibitem{han94} W. Han, K. Pi, W. Bao, K. M. McCreary, Y. Li, W. H. Wang, C. N. Lau, and R. K. Kawakami,
Appl. Phys. Lett. {\bf 94}, 222109 (2009).

\bibitem{han102} W. Han, W. H. Wang, K. Pi, K. M. McCreary, W. Bao, Y. Li, F. Miao, C. N. Lau, and
R. K. Kawakami, Phys. Rev. Lett. {\bf 102}, 137205 (2009).

\bibitem{shiraishi} M. Shiraishi, M. Ohishi, R. Nouchi, N. Mitoma, T. Nozaki, T. Shinjo, and Y. Suzuki,
Adv. Funct. Mater. {\bf 19}, 3711 (2009).


\bibitem{pi} K. Pi, W. Han, K. M. McCreary, A. G. Swartz, Y. Li, and R. K. Kawakami, Phys. Rev. Lett. {\bf 104}, 187201 (2010).

\bibitem{han105} W. Han, K. Pi, K. M. McCreary, Y. Li, J. J. I. Wong, A. G. Swartz, and R. K. Kawakami,
Phys. Rev. Lett. {\bf 105}, 167202 (2010).

\bibitem{yzhou} Y. Zhou and M. W. Wu, Phys. Rev. B {\bf 82}, 085304 (2010).



\bibitem{han107} W. Han and R. K. Kawakami, Phys. Rev. Lett. {\bf 107}, 047207 (2011).

\bibitem{avsar} A. Avsar, T.-Y. Yang, S. Bae, J. Balakrishnan, F. Volmer, M. Jaiswal, Z. Yi, S. R. Ali,
G. G\"{u}ntherodt, B. H. Hong, B. Beschoten, and B. \"{O}zyilmaz, Nano Lett. {\bf 11}, 2363 (2011).

\bibitem{han324} W. Han, K. M. McCreary, K. Pi, W. H. Wang, Y. Li, H. Wen, J. R. Chen, and R. K. Kawakami,
J. Magn. Magn. Mater. {\bf 324}, 369 (2012).

\bibitem{maassen} T. Maassen, I. J. Vera-Marun, M. H. D. Guimar\~{a}es, and B. J. van Wees, Phys. Rev. B
{\bf 86}, 235408 (2012).



\bibitem{wojtaszek} M. Wojtaszek, I. J. Vera-Marun, T. Maassen, 
and B. J. van Wees, Phys. Rev. B {\bf 87}, 081402(R) (2013).

\bibitem{varykhalov} A. Varykhalov, J. S. Barriga, A. M. Shikin, C. Biswas, E. Vescovo, A. Rybkin, D. Marchenko,
and O. Rader, Phys. Rev. Lett. {\bf 101}, 157601 (2008).

\bibitem{neto} A. H. Castro Neto and F. Guinea, Phys. Rev. Lett. {\bf 103}, 026804 (2009).

\bibitem{ertler} C. Ertler, S. Konschuh, M. Gmitra, and J. Fabian, Phys. Rev. B {\bf 80}, 041405(R) (2009).
\bibitem{dugaev} V. K. Dugaev, E. Ya. Sherman, and J. Barna\'{s}, Phys. Rev. B {\bf 83}, 085306 (2011).

\bibitem{pzhang84} P. Zhang and M. W. Wu, Phys. Rev. B {\bf 84}, 045304 (2011).

\bibitem{pzhang14}  P. Zhang and M. W. Wu, New J. Phys. {\bf 14}, 033015 (2012).

\bibitem{morozov} S. V. Morozov, K. S. Novoselov, M. I. Katsnelson, F. Schedin, L. A. Ponomarenko, D. Jiang,
and A. K. Geim, Phys. Rev. Lett. {\bf 97}, 016801 (2006).

\bibitem{jeong} J. S. Jeong, J. Shin, and H. W. Lee, Phys. Rev. B {\bf 84}, 195457 (2011).

\bibitem{pzhang112} P. Zhang, Y. Zhou, and M. W. Wu, J. Appl. Phys. {\bf 112}, 073709 (2012).

\bibitem{dedkov} Y. S. Dedkov, M. Fonin, U. R\"{u}diger, and C. Laubschat, Phys. Rev. Lett. {\bf 100},
107602 (2008).

\bibitem{ryu} S. Ryu, L. Liu, S. Berciaud, Y.-J. 
Yu, H. Liu, P, Kim, G. W. Flynn, and
L. E. Brus, Nano Lett. {\bf 10}, 4944 (2010).

\bibitem{gallo} P. Gallo, A. Arnoult, T. Camps, E. Havard, and C. Fontaine, 
J. Appl. Phys. {\bf 101}, 024322 (2007).


\bibitem{guinea1} F. Guinea, New J. Phys. {\bf 12}, 083063 (2010).

\bibitem{konschuh85} S. Konschuh, M. Gmitra, D. Kochan, and J. Fabian, Phys. Rev. B {\bf 85}, 115423 (2012).

\bibitem{mccann} E. McCann and M. Koshino, Rep. Prog. Phys. {\bf 76}, 056503 (2013).

\bibitem{korm} A. Korm\'{a}nyos and G. Burkard, Phys. Rev. B {\bf 87},
 045419 (2013).

\bibitem{min2} H. Min, B. Sahu, S. K. Banerjee, and A. H. MacDonald, Phys. Rev. B {\bf 75}, 155115 (2007).

\bibitem{castro} E. V. Castro, K. S. Novoselov, S. V. Morozov, N. M. R. Peres, J. M. B. Lopes dos Santos, J. Nilsson,
F. Guinea, A. K. Geim, and A. H. Castro Neto, Phys. Rev. Lett. {\bf 99}, 216802 (2007).

\bibitem{zhang} Y. Zhang, T.-T. Tang, C. Girit, Z. Hao, M. C. Martin, A. 
Zettl, M. F. Crommie, Y. R. Shen, 
and F. Wang, Nature (London) {\bf 459}, 820 (2009).

\bibitem{nilsson} J. Nilsson, A. H. Castro Neto, F. Guinea, and N. M. R. Peres, Phys. Rev. B {\bf 78}, 045405 (2008).

\bibitem{mccann2} E. McCann, arXiv:1205.4849.



\bibitem{viljas} J. K. Viljas and T. T. Heikkil\"{a}, Phys. Rev. B {\bf 81}, 245404 (2010). Its arXiv version arXiv:1002.3502 contains more detailed information.

\bibitem{borysenko} K. M. Borysenko, J. T. Mullen, X. Li, Y. G. Semenov, J. M. Zavada, M. Buongiorno Nardelli, and
K. W. Kim, Phys. Rev. B {\bf 83}, 161402(R) (2011).

\bibitem{cappelluti} E. Cappelluti and G. Profeta, Phys. Rev. B {\bf 85}, 205436 (2012).




\bibitem{yang} T.-Y. Yang, J. Balakrishnan, F. Volmer, A. Avsar, M. Jaiswal, J. Samm, S. R. Ali, A.
Pachoud, M. Zeng, M. Popinciuc, G. G\"{u}ntherodt, B. Beschoten, and 
B. \"{O}zyilmaz, Phys. Rev. Lett. {\bf 107}, 047206 (2011).





\bibitem{diez} M. Diez and G. Burkard, Phys. Rev. B {\bf 85}, 195412 (2012).

\bibitem{neumann} I. Neumann, J. Van de Vondel, G. Bridoux, M. V. Costache, F. Alzina, C. M.
S. Torres, and S. O. Valenzuela, Small {\bf 9}, 156 (2013).

\bibitem{dp} M. I. D'yakonov and V. I. Perel', Zh. Eksp. Teor. Fiz. {\bf 60}, 1954 (1971)
[Sov. Phys. JETP {\bf 33}, 1053 (1971)].


\bibitem{wureview} M. W. Wu, J. H. Jiang, and M. Q. Weng, Phys. Rep. {\bf
    493}, 61 (2010).

\bibitem{wuning} M. W. Wu and C. Z. Ning, Eur. Phys. J. B {\bf 18}, 373
  (2000); M. W. Wu, J. Phys. Soc. Jpn. {\bf 70}, 2195 (2001).




\bibitem{weng1} M. Q. Weng and M. W. Wu, Phys. Rev. B {\bf 68}, 075312 (2003);
{\bf 70}, 195318 (2004).


\bibitem{glazov} M. M. Glazov and E. L. Ivchenko, Pis'ma Zh. Eksp. Teor. Fiz. {\bf 75}, 476 (2002);
Zh. Eksp. Teor. Fiz. {\bf 126}, 1465 (2004) [JETP Lett. {\bf 75}, 403 (2002); JETP {\bf 99}, 1279 (2004)].

\bibitem{jzhou} J. Zhou, J. L. Cheng, and M. W. Wu, Phys. Rev. B {\bf
    75}, 045305 (2007).

\bibitem{leyland} W. J. H. Leyland, G. H. John, R. T. Harley,
  M. M. Glazov, E. L. Ivchenko, D. A. Ritchie, I. Farrer,
  A. J. Shields, and M. Henini, Phys. Rev. B {\bf 75}, 165309 (2007).

\bibitem{stich} D. Stich, J. Zhou, T. Korn, R. Schulz, D. Schuh, W. Wegscheider, M. W. Wu, and
C. Sch\"{u}ller, Phys. Rev. Lett. {\bf 98}, 176401 (2007); Phys. Rev. B {\bf 76}, 205301 (2007).



\bibitem{lfhan43} L. F. Han, X. H. Zhang, H. Q. Ni, and Z. C. Niu,
  Physica E {\bf 43}, 1127 (2011).



\bibitem{koster} G. F. Koster, J. O. Dimmock, R. G. Wheeler, and H. Statz, {\it Properties of the Thirty-two Point Groups}
(MIT Press, Cambridge, Massachusetts, 1963).


\bibitem{jiang} J. H. Jiang and M. W. Wu, Phys. Rev. B {\bf 79}, 125206 (2009).

\bibitem{fzhang} F. Zhang, H. Z. Zheng, Y. Ji, J. Liu, and G. R. Li, 
Europhys. Lett. {\bf 83}, 47006 (2008).

\bibitem{korn} T. Korn, D. Stich, R. Schulz, D. Schuh, W. Wegscheider, 
and C. Sch\"{u}ller, Adv. Solid State Phys. {\bf 48}, 143 (2009).


\bibitem{delta} It is noted that when dealing with the delta function in the scattering terms,
we neglect the anisotropy of the energy spectrum since the effect of the anisotropy on spin
relaxation is marginal.\cite{diez}


\bibitem{piscanec} S. Piscanec, M. Lazzeri, F. Mauri, A. C. Ferrari, and J. Robertson, Phys. Rev. Lett.
{\bf 93}, 185503 (2004).

\bibitem{lazzeri} M. Lazzeri, S. Piscanec, F. Mauri, A. C. Ferrari, and J. Robertson, Phys. Rev. Lett.
{\bf 95}, 236802 (2005).


\bibitem{rana} F. Rana, P. A. George, J. H. Strait, J. Dawlaty, S. Shivaraman, M. Chandrashekhar, and M.
G. Spencer, Phys. Rev. B {\bf 79}, 115447 (2009).

\bibitem{peskin} M. E. Peskin and D. V.
Schroeder, {\it An Introduction to Quantum Field Theory} 
(Addison-Wesley, New York, 1995).

\bibitem{divincenzo} D. P. DiVincenzo and E. J. Mele, Phys. Rev. B {\bf 29}, 1685 (1984).

\bibitem{dresselhaus} G. Dresselhaus, Phys. Rev. {\bf 100}, 580 (1955).

\bibitem{rashba} Y. A. Bychkov and E. I. Rashba, J. Phys. C {\bf 17}, 6039 
(1984); Pis'ma Zh. Eksp. Teor. Fiz {\bf 39}, 66 (1984) [JETP Lett. {\bf 39}, 78 (1984)].

\bibitem{ed} The electron density in the experiment by Han and 
Kawakami\cite{han107} is obtained from its arXiv version (arXiv:1012.3435).

\bibitem{sdc} It is noted that due to the complicated energy spectrum of BLG, we employ the single-layer
graphene approximation to calculate the spin diffusion coefficient in BLG
according to the report by Zhang and Wu.\cite{pzhang84}



\bibitem{sarma81} S. Das Sarma, E. H. Hwang, and E. Rossi, 
Phys. Rev. B {\bf 81}, 161407(R) (2010).

\bibitem{adam82} S. Adam and M. D. Stiles, Phys. Rev. B {\bf 82}, 075423 (2010).

\bibitem{pade} G. A. Baker, Jr., {\it Essentials of Pad\'{e} Approximants} 
(Academic, New York, 1975).


\bibitem{mahan} G. D. Mahan, {\it Many-Particle Physics} (Plenum, New York, 1990).

\bibitem{adam1} S. Adam, E. H. Hwang, V. M. Galitski, and S. Das Sarma, Proc. Natl. Acad. Sci. U.S.A. {\bf 104},
18392 (2007).

\bibitem{hwang} E. H. Hwang, S. Adam, and S. Das Sarma, Phys. Rev. Lett. {\bf 98}, 186806 (2007).

\bibitem{adam2} S. Adam and S. Das Sarma, Solid State Commun. {\bf 146}, 356 (2008).

\bibitem{wunsch} B. Wunsch, T. Stauber, F. Sols, and F. Guinea, New J. Phys. {\bf 8}, 318 (2006).

\bibitem{xfwang} X.-F. Wang and T. Chakraborty, Phys. Rev. B {\bf 75}, 033408 (2007).

\bibitem{hwang75} E. H. Hwang and S. Das Sarma, Phys. Rev. B {\bf 75}, 205418 (2007).


\bibitem{fratini} S. Fratini and F. Guinea, Phys. Rev. B {\bf 77}, 195415 (2008).

\bibitem{gamayun} O. V. Gmayun, Phys. Rev. B {\bf 84}, 085112 (2011).



\bibitem{hwang2} E. H. Hwang and S. Das Sarma, Phys. Rev. B {\bf 77}, 115449 (2008).

\bibitem{jhchen} J.-H. Chen, C. Jang, S. Xiao, M. Ishigami, and M. S. Fuhrer, Nat. Nanotechnol. {\bf 3},
206 (2008).








\end{thebibliography}
\end{document}